\documentclass[transmag]{IEEEtran}
\usepackage{balance}
\usepackage{cite}
\usepackage{amsmath,amssymb,amsfonts}
\usepackage{textcomp}
\usepackage{algorithm}
\usepackage[normalem]{ulem}
\usepackage{array}
\usepackage{float}
\usepackage{longtable}
\usepackage{makecell}
\usepackage{placeins}
\usepackage[pdftex]{graphicx}
\usepackage{graphicx}
\DeclareGraphicsExtensions{.pdf,.jpeg,.png,.eps}
\usepackage{multirow}
\usepackage{mathtools}
\usepackage{algpseudocode}
\usepackage{supertabular}
\usepackage{cite}
\usepackage[nodisplayskipstretch]{setspace}
\usepackage{color}
\def\BibTeX{{\rm B\kern-.05em{\sc i\kern-.025em b}\kern-.08em T\kern-.1667em\lower.7ex\hbox{E}\kern-.125emX}}
\begin{document}

\title{Multiple Access in Aerial Networks: From Orthogonal and Non-Orthogonal to Rate-Splitting}

\author{Wael Jaafar, \IEEEmembership{Senior Member, IEEE}, Shimaa Naser, Sami Muhaidat, \IEEEmembership{Senior Member, IEEE}, \\ Paschalis C. Sofotasios,
\IEEEmembership{Senior Member, IEEE}, and Halim Yanikomeroglu,
\IEEEmembership{Fellow, IEEE}
\thanks{W. Jaafar and  H. Yanikomeroglu are with the Department of Systems and Computer Engineering, Carleton University, Ottawa,
ON, Canada, (e-mails: \{waeljaafar, halim\}@sce.carleton.ca.)}
\thanks{S. Naser and S. Muhaidat are with the Center for Cyber-Physical Systems, Department of Electrical Engineering and Computer Science, Khalifa University, Abu Dhabi, UAE, (e-mails: 100049402@ku.ac.ae; muhaidat@ieee.org.)}
\thanks{P. C. Sofotasios is with the Center for Cyber-Physical Systems, Department of Electrical Engineering and Computer Science, Khalifa University, Abu Dhabi, UAE, and also with the Department of Electrical Engineering, Tampere University, Tampere, Finland, (e-mail: p.sofotasios@ieee.org.)}
\thanks{This work is supported in part by the Natural Sciences and Engineering Research Council Canada (NSERC) and in part by Khalifa University under Grant
No. KU/RC1-C2PS-T2/8474000137 and Grant No. KU/FSU-474000122.}
}

\IEEEtitleabstractindextext{\begin{abstract}
Recently, interest on the utilization of unmanned aerial vehicles (UAVs) has aroused. Specifically, UAVs can be used in cellular networks as aerial users for delivery, surveillance, rescue search, or as an aerial base station (aBS) for communication with ground users in remote uncovered areas or in dense environments requiring prompt high capacity. Aiming to satisfy the high requirements of wireless aerial networks, several multiple access techniques have been investigated. In particular, space-division multiple access (SDMA) and power-domain non-orthogonal multiple access (NOMA) present promising multiplexing gains for aerial downlink and uplink. Nevertheless, these gains are limited as they depend on the conditions of the environment. Hence, a generalized scheme has been recently proposed, called rate-splitting multiple access (RSMA), which is capable of achieving better spectral efficiency gains compared to SDMA and NOMA. In this paper, we present a comprehensive survey of key multiple access technologies adopted for aerial networks, where aBSs are deployed to serve ground users. Since there have been only sporadic results reported on the use of RSMA in aerial systems, we aim to extend the discussion on this topic by modelling and analyzing the weighted sum-rate performance of a two-user network served by an RSMA-based aBS. Finally, related open issues and future research directions are exposed.
\end{abstract}

\begin{IEEEkeywords}
Orthogonal multiple access (OMA), non-orthogonal multiple access (NOMA), rate-splitting multiple access (RSMA), unmanned aerial vehicle (UAV), survey.
\end{IEEEkeywords}
}

\maketitle

\section{INTRODUCTION}

\IEEEPARstart{D}{ue} to the unprecedented growth of mobile data traffic and stringent quality-of-service (QoS) requirements, recent research efforts have focused on several key enabling technologies for 5G networks and beyond \textcolor{black}{\cite{Zhang2020_o}}, such as millimeter waves (mmWave), terahertz (THz) communications, {multiple-input multiple-output} (MIMO), massive MIMO, multiple access techniques, relaying, cognitive radio, and unmanned aerial networks \textcolor{black}{\cite{Liu2019_o}}. In particular, unmanned aerial vehicles (UAVs) have demonstrated great potential in enabling new applications. For instance, UAVs can be used for aerial security inspection, traffic monitoring, smart agriculture, aerial delivery, etc. {\cite{Li2019,Zeng2019} (and references therein)}.
Furthermore, UAVs can be deployed as aerial base stations (aBSs) to provide wireless access to ground {and aerial} devices/users, in several scenarios 
such as temporary events, disasters when a terrestrial cellular network is not fully operational, and congestion due to unpredictable traffic surges, as well as aerial devices/users (e.g., cargo drones) \cite{Cao2018}, {\cite{cherif2020optimal}}.

Unlike conventional cellular networks, UAVs exploit their additional degrees of freedom, provided by their own 3D mobility, agility, and high line-of-sight (LoS) probability, to offer strong non-terrestrial communication links to mobile users with versatile quality of service requirements \cite{Alzenad2018,3GPP38811}. Nevertheless, these additional degrees of freedom cannot provide the best communication experience to users of large-scale cellular networks. Meanwhile, downlink multiple access techniques have played a crucial role in achieving improved system performance in terms of data rate, outage probability, and latency, in a resource-scarce environment. More specifically, orthogonal and non-orthogonal multiple access schemes (OMA and NOMA) have been proposed for aerial networks, including {}{time-division multiple access} (TDMA) \cite{Lyu2016,Yin2019}, {}{frequency-division multiple access} (FDMA) \cite{Yin2019}, orthogonal FDMA (OFDMA) \cite{Wu2018}, {}{code-division multiple access} (CDMA) \cite{Ho2010}, and {}{space-division multiple access} (SDMA) \cite{Jiang2012} as OMA techniques. In TDMA, different time slots are allocated to different users communicating with a UAV. In FDMA and OFDMA, different frequency bands or subcarriers are assigned to users, whereas in CDMA, by contrast, orthogonal codes are assigned to different users. Finally, in SDMA, spatial separation between groups of users using beamforming in order to provide each group with full time or frequency resources \cite{Lyu2016,Yin2019,Wu2018,Ho2010,Jiang2012}. 

In the same context, NOMA has been identified as a key enabler for 5G and beyond-5G due to its superior spectrum efficiency.
In particular, NOMA was introduced as a study item in the 3rd Generation Partnership Project (3GPP) New Radio (NR) Release 16 \cite{Liu2017,3GPP38812}. The key concept of NOMA rests on assigning either different power levels to users based on their channel gains, known as power-domain NOMA (P-NOMA) \cite{Islam2017,maraqa2019survey}, or on different codes, referred to as code-domain NOMA (C-NOMA) \cite{Cai2018}. The overwhelming literature has mainly focused on the integration of P-NOMA into UAV networks \cite{Li2019,Liu2019}. P-NOMA relies on superposition coding at the transmitter and successive interference cancellation (SIC) at the receiver in order to differentiate between involved signals. It is worth noting that P-NOMA can be used by a single-antenna or multi-antenna aBS \cite{Ding2014,Liu2018_3}. In the first setup, i.e., single-antenna, only the power allocation challenge has to be solved. However, in the second setup, more complex precoding schemes need to be \textcolor{black}{considered}. 

However, despite its promising performance, MIMO NOMA may not be optimal for single-antenna multi-user (MU-MISO) systems. In fact, MU-MISO NOMA performs well in overloaded conditions, i.e., where the number of users to serve exceeds the number of transmit antennas, but at the cost of increased receiver complexity. However, in underloaded conditions, SDMA has been shown to be a more appropriate option. 

Recently, rate-splitting multiple access (RSMA) was proposed as a powerful and generalized technique for MIMO-based multiple access systems {\cite{Clerckx2016,Mao2018,Mao2018_3,dizdar2020ratesplitting}}. RSMA is expected to be one of the leading beyond-5G technologies, as it has been already shown to outperform SDMA and NOMA in underloaded and overloaded networks, respectively. Furthermore, its integration into aerial networks will provide enhanced performance in terms of {}{spectral efficiency} (SE) and energy efficiency (EE). 
\textcolor{black}{It is worth mentioning that RSMA technique can be used in uplink and downlink, but with different motivations and structures. Downlink RSMA aims to retain the benefits of SDMA and NOMA, namely high spectral efficiency, high energy efficiency, and low or null interference, while addressing their limitations, such as transceiver complexity, and sensitivity to imperfect channel knowledge at the transmitter and channel strengths \cite{Mao2018}, whereas uplink RSMA targets achieving the capacity region without requiring any synchronization/coordination among users \cite{Rimoldi1996}.}

The open technical literature on multiple access techniques has greatly matured, particularly concerning OMA and NOMA techniques, \textcolor{black}{with an interesting number of survey papers  \cite{Nguyen1990,Sternad2007,Akka2011,Wang2012,Agiwal2016,Liu2017,Islam2017,Ding2017,Dai2018,Cai2018,Vaezi2019,maraqa2019survey,Makki2020}. However, {little attention has been given to} multiple access techniques in the context of aerial networks \cite{Xiao2016,maraqa2019survey,Li2019,Zhang2019surv,Zeng2019}}. {The interest of multiple access in aerial systems is motivated by several applications.
Of particular interest is a UAV acting as aBS by serving ground devices, providing ubiquitous connectivity to support the seamless integration of  heterogeneous networks. In this context, multiple access techniques are expected to provide sufficient resources and satisfy the requirements of served users. As aerial users or relays, UAVs utilize multiple access techniques to offload efficiently the generated or transiting traffic. This can be of great importance in applications such as sensitive Internet-of-things (IoT) data collection and processing, cellular coverage extension, etc.}

Moreover, as RSMA is attracting the attention of the research community, to our best knowledge, there has not been any survey which summarizes the up-to-date work. Motivated by the above, in this contribution we shed light on some of the most efficient multiple access schemes for aerial networks, taking into account that MU-MISO multiple access techniques for aBSs are still in their infancy. Specifically, we present a comprehensive study of OMA and NOMA techniques, \textcolor{black}{using either SISO or MISO channels}, with particular attention to aerial systems\footnote{\textcolor{black}{It is worth noting that this survey covers both SISO and MISO channels setups since, depending on the targeted application, e.g., communications in cellular networks, IoT-based data collection, edge computing, etc, one of these channel types can be considered.}}. Then, we overview current RSMA literature, and address its integration into aerial systems. Finally, open issues and some interesting research directions are discussed. To the best of our knowledge, this is the first comprehensive survey that covers all of existing multiple access techniques for aerial networks. Table \ref{TableSurvey} below presents a comprehensive summary of surveys that covered multiple access techniques and/or aerial networks.

\begin{table*}
\centering
\caption{Summary of relevant survey papers.}
\label{TableSurvey}
\footnotesize
\begin{tabular}{|p{65pt}|p{20pt}|p{85pt}|p{170pt}|p{50pt}|p{50pt}|}
\hline
\makecell{\textbf{Survey paper}} & \makecell{\textbf{Year}} & \makecell{\textbf{Topic}} & \makecell{\textbf{Addressed issues}} & \makecell{\textbf{Multiple}\\ \textbf{access}} & \makecell{\textbf{Aerial}\\ \textbf{networks}}  \\
\hline 
\hline

\makecell{Agiwal \textit{et al.} \cite{Agiwal2016}} & \makecell{2016} & \makecell{5G enabling \\technologies.} & \makecell{5G architecture, mmWave, massive MIMO, \\full-duplex, beamforming, multiple access, \\applications, quality expectations, sustainability, \\and related challenges.} & \makecell{SDMA.} & \makecell{Not covered} \\ \hline

\makecell{Xiao \textit{et al.} \cite{Xiao2016}} & \makecell{2016} & \makecell{mmWave-based aerial \\networks.} & \makecell{mmWave channel propagation, fast beamforming \\design and tracking, SDMA-based mmWave system, \\blockage issues, and user discovery.} & \makecell{SDMA.} & \makecell{Covered} \\ \hline



\makecell{Islam \textit{et al.} \cite{Islam2017}} & \makecell{2017} & \makecell{P-NOMA in 5G systems.} & \makecell{Basics of NOMA, cooperative NOMA, \\MIMO-NOMA, NOMA in visible light \\communications, NOMA performances, NOMA \\challenges, and implementation issues.} & \makecell{P-NOMA.} & \makecell{Not covered} \\ \hline

\makecell{Cai \textit{et al.} \cite{Cai2018}} & \makecell{2018} & \makecell{Modulation and multiple \\access techniques \\for 5G.} & \makecell{Modulation techniques in OMA, C-NOMA, \\ P-NOMA, multi-antenna NOMA, \\cooperative NOMA, and NOMA multiplexing.} & \makecell{OMA, \\ and NOMA.} & \makecell{Not covered} \\ \hline

\makecell{Maraqa \textit{et al.} \cite{maraqa2019survey}} & \makecell{2019} & \makecell{Integration of P-NOMA \\in future technologies of \\wireless networks.} & \makecell{P-NOMA, conventional and future \\wireless technologies, such as massive MIMO, \\ mmWave, coordinated multi-point, cooperation, \\UAV-assisted communications, etc., \\and future research directions. } & \makecell{P-NOMA.} & \makecell{Covered} \\ \hline

\makecell{Li \textit{et al.} \cite{Li2019}} & \makecell{2019} & \makecell{Advances and trends in \\aerial communications \\for 5G and beyond.} & \makecell{Space-air-ground integrated networks, mmWave \\UAV, UAV NOMA, cognitive UAV networks, \\energy harvesting UAV networks, UAV-assisted \\Hetnets, combined UAV and device-to-device \\networks, software-defined UAV networks, \\UAV-based edge computing, \\caching in UAV.} & \makecell{P-NOMA.} & \makecell{Covered} \\ \hline

\makecell{Zhang \textit{et al.} \cite{Zhang2019surv}} & \makecell{2019} & \makecell{mmWave UAV-assisted \\wireless networks.} & \makecell{mmWave antenna technique, radio propagation \\channel, multiple access, UAV spatial configuration, \\resource optimization, security, and performance.} & \makecell{P-NOMA\\ and SDMA. } & \makecell{Covered} \\ \hline

\makecell{Zeng \textit{et al.} \cite{Zeng2019}} & \makecell{2019} & \makecell{UAV communication \\technologies for 5G \\and beyond.} & \makecell{UAV-assisted wireless communication performance \\analysis and parameters optimization, \\ cellular-connected UAV experiments \\and performance,  UAV swarm communication, \\security, mmWave, caching, and edge computing.} & \makecell{P-NOMA.} & \makecell{Covered} \\ \hline

\makecell{This survey} & \makecell{2020} & \makecell{{Multiple access in aerial} \\{networks}.} & \makecell{UAV communication background, conventional \\ {OMA and NOMA} for aerial networks, \\ RSMA communication background, and {integration} \\ {of RSMA into aerial networks}.} & \makecell{{OMA, NOMA}, \\ and {RSMA.}} & \makecell{Covered} \\ \hline

\end{tabular}
\end{table*}



\begin{figure}[t]
\centering
\includegraphics[width=0.98\columnwidth]{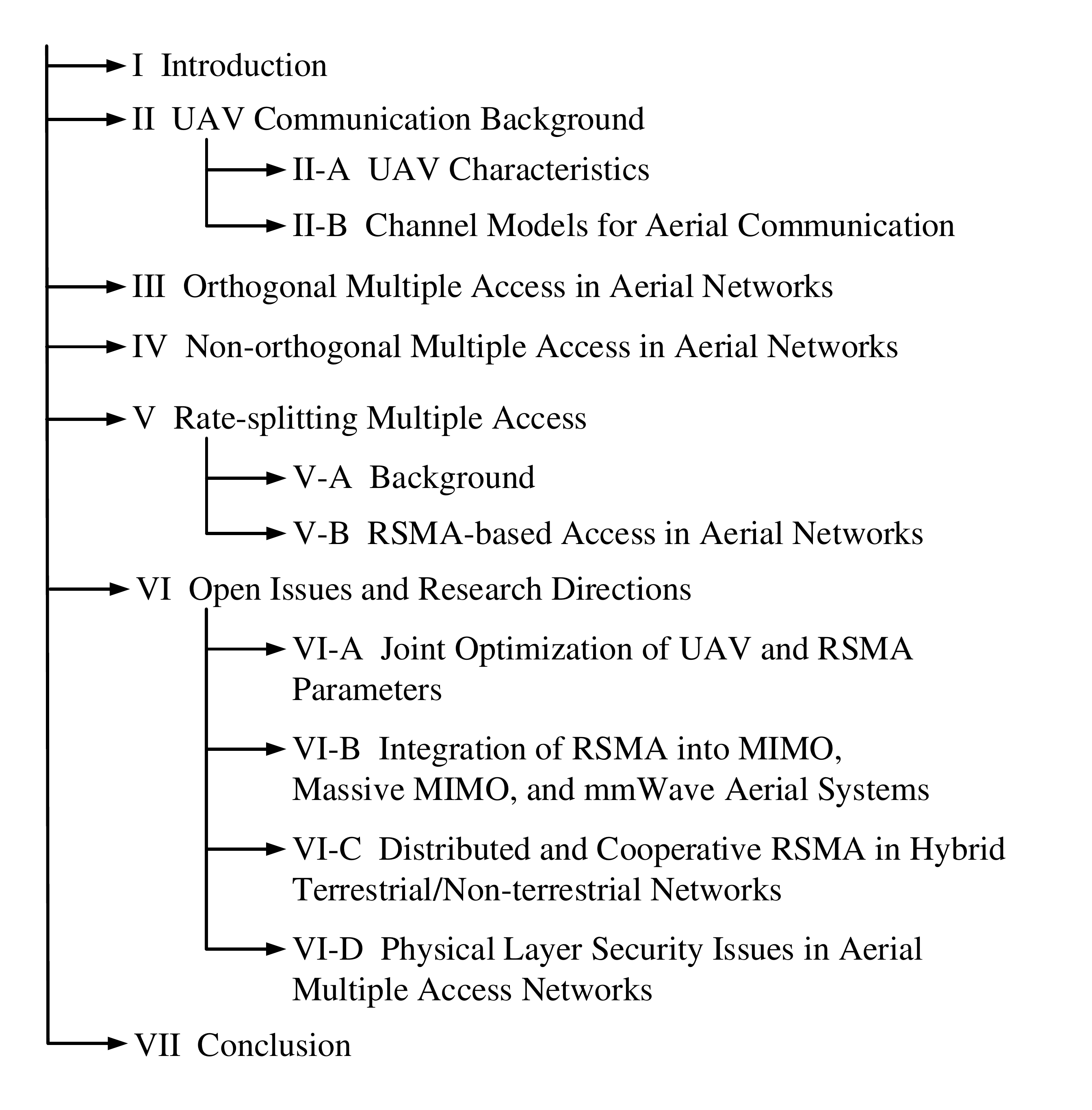}
\caption{Organization of the present work.}
\label{Fig:organization}
\end{figure}

The rest of the paper is organized as illustrated in Fig. \ref{Fig:organization}. Section II presents an overview of UAV communication characteristics. Sections III and IV cover OMA and NOMA techniques for aerial networks, respectively. In Section V, the recently proposed RSMA technique is presented, then analyzed for a multi-user system served by an aBS. Section VI discusses open issues and research directions that would unleash RSMA's potential gains for aerial networks. Finally, Section VII concludes the paper.   

\begin{table}[t]
\scriptsize
\centering
\caption{List of Abbreviations.}
\begin{tabular} [c]{p{2cm} p{5cm} }
aBS & Aerial base station \\
AO	&Alternating optimization\\ 
AS & Airborne station \\
CDMA &	{}{Code-division multiple access}\\ 
CD-NOMA& Code-domain NOMA\\ 
C-RAN & Cloud radio access network \\
CSIT 	& Channel state information at transmitter\\ 
EE & {}{Energy efficiency} \\
{ESR} & {Ergodic sum rate} \\
FDMA & {Frequency-division multiple access} \\
gBS & Ground base station \\
gUE & Ground user \\
HRS & {}{Hierarchical rate-splitting} \\
LoS 	&Line-of-sight\\ 
MIMO &	{}{Multiple-input multiple-output}\\ 
MISO	&{}{Multiple-input single-output}\\ 
ML & {}{Machine learning} \\
MSE 	&Mean-square error\\ 
NLoS & {Non-line-of-sight} \\
NOMA &	Non-orthogonal multiple access\\ 
OFDMA	& {}{Orthogonal frequency-division multiple access}\\ 
OMA & Orthogonal multiple access \\
P-NOMA& 	Power-domain NOMA\\ 
QoS 	&Quality-of-service\\ 
RS & Rate-splitting \\
RSMA	&Rate-splitting multiple access\\ 
SDMA	& {}{Space-division multiple access}\\ 
SE & {}{Spectral efficiency} \\
SIC 	&Successive interference cancellation\\ 
{}{SINR} & {}{Signal-to-interference-plus-noise ratio}\\
\textcolor{black}{SNR} & \textcolor{black}{Signal-to-noise ratio}\\
SWAP & Size, weight, and power \\
{}{SWIPT} & {}{Simultaneous wireless information and power transfer}\\
TDMA &	{}{Time-division multiple access}\\ 
UAV & Unmanned aerial vehicle \\
WMMSE	&Weighted minimum mean-square error\\
WSR	&Weighted sum rate \\ 
ZF & Zero-forcing \\
\end{tabular}
\label{TableI}
\end{table}

\section{UAV Communication Background}\label{sec:background}
In this section, 
we present UAV characteristics and communication channel models. Used abbreviations are summarized in Table \ref{TableI}.

\subsection{UAV Characteristics}
With the technological advancements and cost reduction in UAV manufacturing and communication equipment miniaturization, it is now possible to build small-size aBSs and deploy them to serve ground users. Unlike terrestrial networks, where topology is static and BSs are fixed, UAVs bring flexibility in many aspects {}{\cite{Zeng2016_2}}. Accordingly, standardization efforts are accelerating in order to include 5G enhancements for UAVs in 3GPP NR Release 17 \cite{3GPPUAV}.

Since aBSs can be deployed on demand {}{\cite{Sharma2016_2}}, they are more likely to experience better LoS links to ground users than gBSs, owing to their high altitude. Furthermore, their 3D mobility provides an additional degree of freedom to improve communication performance compared to gBSs.      
Nevertheless, the exploitation of these advantages to enhance performance poses several challenges, including UAV's stringent size, weight, and power (SWAP) constraints, as well as the joint design of UAV mobility control and communication scheduling and/or resource allocation. In the following, we elaborate on these points in further details.
\begin{enumerate}
    \item \textit{High altitude:} UAVs can be positioned at higher altitudes than gBSs. The latter have a maximum altitude of 25 m in urban environments \cite{3GPP36777}, whereas UAVs are allowed to fly up to {}{122 m (400 ft)} \cite{FAA2019}. At these higher altitudes, aBSs can achieve a wider ground coverage compared to gBSs. It is worth noting that this altitude limit is for toy-type UAVs. It is possible that, when operators start utilizing aBSs, they will likely be allowed to operate them at altitudes above {}{122 m} when necessary.
    \item \textit{High LoS probability:} Terrestrial channels between gBSs and ground users suffer from severe path-loss due to shadowing and multi-path fading. However, aBSs, positioned at higher altitudes, experience less of these channel effects, and hence benefit from dominant LoS links with a higher probability. As such, better communication performance can be achieved, and more flexibility is obtained for aBS-user association. However, these channels may cause significant co-channel interference to coexisting gBS/aBS served users. Consequently, interference mitigation is a crucial challenge that must be addressed in aerial networks.   
    \item \textit{High 3D mobility:} aBSs enjoy mobility that gBSs lack. With this key advantage, aBSs can move at high speeds to serve both static and dynamically moving users. This mobility, however, may cause frequent handovers and varying backhaul links that require an appropriate UAV mobility design. For instance, 3D placements and trajectories of aBSs must be designed to adapt to the mobility and channel dynamics of ground users, while continuously satisfying their QoS requirements. To further enhance performance, this operation can be jointly considered and addressed with appropriate communication protocols, such as scheduling and resource allocation {}{\cite{Alzenad2018,Li2019_2}}.
  
    \item \textit{SWAP constraints:} Unlike gBSs, which benefit from stable power supplies or rechargeable batteries, SWAP constraints represent major challenges for UAVs in achieving sustainable and long-term operations. This is mainly due to the fact that aBSs are expected to be small-sized and light-weight in order to sustain longer operation times. Moreover, UAV mobility dynamics, i.e., flying and hovering, are energy-consuming compared to communication. Consequently, energy-efficient UAV designs will achieve significant operation improvements for aBSs \cite{Zeng2017,Galkin2018UAVsAM}.   
\end{enumerate}

\subsection{Channel Models for Aerial Communication}
In this subsection, we present the mathematical channel models for communication between an aBS and aerial or ground users. Channels between aBSs and aerial users are accurately characterized by the simple free-space path-loss model \cite{Ahmed2016}, where the channel gain can be described as 
\begin{equation}
    \label{eq:PL}
    h=\sqrt{\beta(d)}=\sqrt{\left( \frac{\lambda}{4 \pi d} \right)^2}=\sqrt{\beta_0 d^{-2}},
\end{equation}
where $d$ represents the distance between an aBS and its user, $\lambda$ is the carrier wavelength, and $\beta_0=\frac{\lambda}{4 \pi}$ is the channel power at reference distance $d=1$ m.
It should be noted that different types of large-scale and small-scale channel models have been considered for the aBS-ground user (aBS-gUE) channel. According to \cite{Zeng2019}, large-scale aBS-gUE channel models have a tripartite classification: a \textit{free-space channel model}; \textit{altitude/angle-dependent models}; and a \textit{probabilistic LoS channel model}. 

For the \textit{free-space channel model}, the same channel expression (\ref{eq:PL}) can be considered for aBS-gUE links. It is a reasonable model for practical rural area environments, and/or when the aBS is flying at a sufficiently high altitude to achieve a clear LoS link. Nevertheless, this model is oversimplified in urban and dense-urban environments or for low-altitude aBSs.  

In urban environments, as a UAV ascends, the effects of signal obstruction and scattering are reduced. \textit{Altitude/angle-dependent models} take this into account by adjusting altitude- or angle-dependent parameters, such as path-loss exponent, the Rician K-factor, or the variance of random shadowing \cite{Amorim2017} and \cite{Azari2018}. {}{For instance, the path-loss exponent, denoted $\alpha$, is modelled as a monotonically decreasing function of the UAV altitude in \cite{Amorim2017}
\begin{equation}
\label{eq:beta}
    \alpha= \max \left( p_1 - p_2 \; \text{log}_{10}(q), 2 \right),
\end{equation}
where $q$ is the UAV's altitude, and $p_1$ and $p_2$ are modelling parameters obtained via curve fitting of measured channel results. Eq. (\ref{eq:beta}) emphasizes that as $q$ increases, less obstacles and scattering exist, and subsequently path-loss exponent becomes smaller.
Nevertheless, this altitude-dependent channel model fails to model the environment changes at a fixed $q$. To address this, the works in \cite{Azari2018} takes into account the elevation angle, denoted $\theta$, between the aBS and gUE, which depends on both $q$ and the horizontal distance between aBS and gUE (i.e., distance between the projection of aBS on the ground and gUE). As an example, for the considered aBS-gUE Rician fading channel in \cite{Azari2018}, the Rician K-factor and $\alpha$ are modelled as a non-decreasing and non-increasing functions of $\theta$, respectively. That means, as $\theta$ increases (i.e., the aBS flies either higher or closer to the gUE), the LoS component becomes dominant.     
}

Another accurate model for urban and dense-urban environments is the \textit{probabilistic LoS channel model}. This model distinguishes between LoS and {NLoS} propagation by taking their occurrence probabilities into account \cite{Hourani2014}. The channel coefficients can be expressed by
\begin{equation}
    \label{eq:Hourani}
    h=\sqrt{\beta_{\rm{LoS}}(d)P_{\rm{LoS}}+\beta_{\rm{NLoS}}\left(1-P_{\rm{LoS}}\right)},
\end{equation}
\textcolor{black}{where $\beta_{\rm{LoS}}=\beta_0 d^{-\alpha}$, $\beta_{\rm{NLoS}}=\kappa \beta_0 d^{-\alpha}$, $\alpha \in [2,6]$ is the path-loss exponent}, $\kappa < 1$ is the additional attenuation factor due to NLoS, and $P_{\rm{LoS}}$ is the probability of the presence of a LoS link, {}{modelled as a logarithmic function of the elevation angle $\theta$ and} given by
\begin{equation}
    \label{eq:PLoS}
    P_{\rm{LoS}}=\frac{1}{1+a \; e^{-b(\theta-a)}},
\end{equation}
with $a$ and $b$ the modeling parameters, obtained from measurements \cite{Hourani2014}.

\begin{figure*}[t]
\centering
\includegraphics[width=0.95\linewidth]{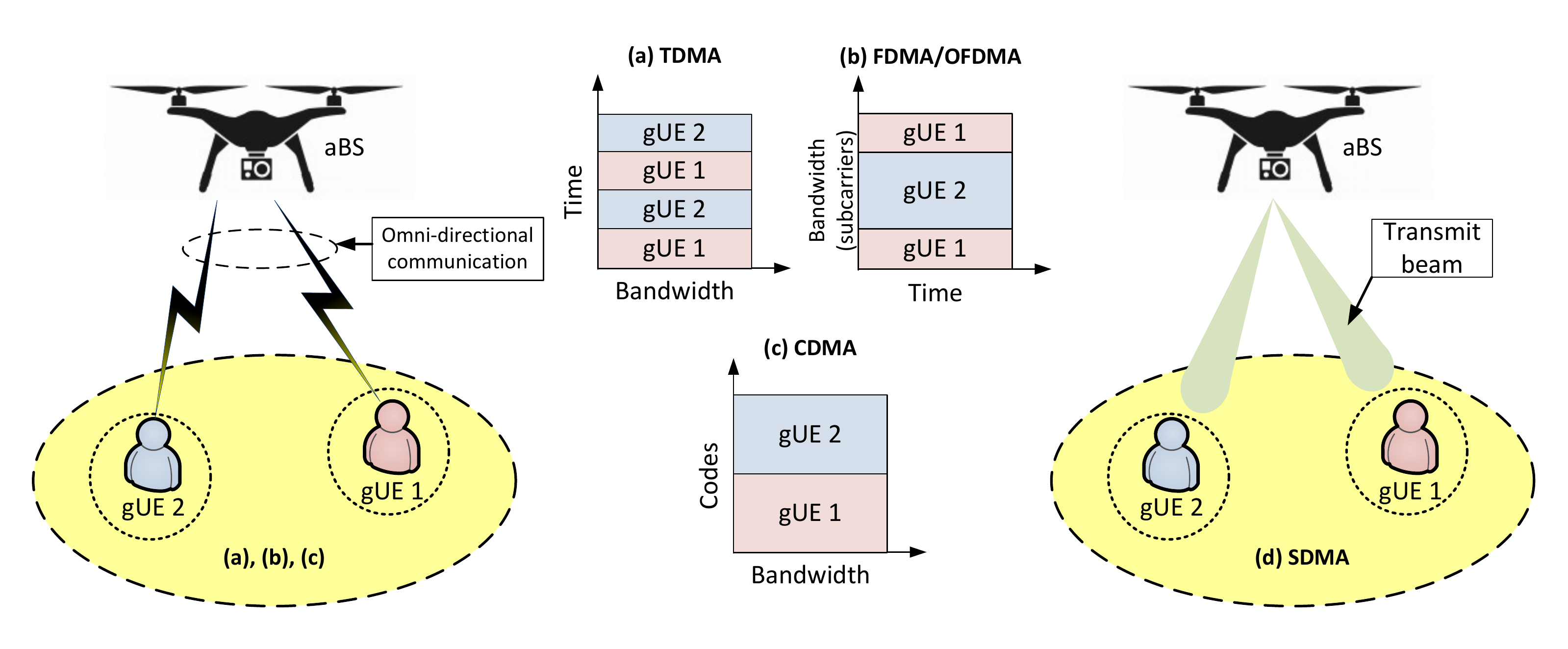}
\caption{Illustration of OMA techniques in a two-user aerial system: (a) TDMA (b) FDMA/OFDMA (c) CDMA (d) SDMA.}
\label{Fig:OMA}
\end{figure*}

In \cite{Khuwaja2018}, different types of  aBS-gUE small-scale fading were presented, i.e., the Loo model, Rayleigh model, Nakagami-$m$ model, Rician model, and Werbull model. It was shown that the Rician model is an adequate generalized small-scale model that can be considered for dense-urban, urban, and suburban environments \cite{Holzbock1999,Newhall2003,Simunek2013,Ye2017}. Within this context, the small-scale fading coefficients of a Rician channel are given by
\begin{equation}
    \label{eq:smallscale}
    g=\sqrt{\frac{K}{K+1}}\bar{g}+\sqrt{\frac{1}{K+1}}\tilde{g},
\end{equation}
\textcolor{black}{where $K$ is the Rician K-factor expressed as $K(\theta)=a_1 e^{b_1 (\theta- a_1) }$, such that $a_1=K(0)$ and $b_1=\frac{2}{\pi}\text{ln}\left(\frac{K(\frac{\pi}{2})}{K(0)} \right)$ are constants that depend on the environment and system parameters \cite{Azari2018}, $\bar{g}$ is the deterministic LoS component with $|\bar{g}|=1$, and $\tilde{g}$ is the random NLoS component.} The latter is modeled by a complex Gaussian random variable with zero mean and unit variance and is denoted by $\mathcal{CN}(0,1)$. Finally, the generalized channel model, denoted by $f$, can be written as $f=h \cdot g$.

Most of these models can be extended to a MIMO scenario. However, doing this fails to consider the particularities of UAVs, such as high altitudes and mobility. In fact, MIMO-based UAV channel modeling is still in its infancy and only a few works have taken interest in this issue. For instance, \cite{Jiang2012} assumed a correlated Rician fading channel between single-antenna users and a multi-antenna UAV, while considering large-scale path loss. The authors of \cite{Zeng2017_4} studied the space-time correlation function of a 3D mobile MIMO-based UAV, where only the NLoS channel was considered. In \cite{Jiang2018}, a 3D geometric MIMO channel model was developed for the aBS-gUE link, where the elliptic cylinder model was adopted to describe the distribution of scatters around the aBS and gUE.

\section{Orthogonal Multiple Access in Aerial Networks}
Inspired by the promising multiplexing gains of multiple access techniques in terrestrial networks, researchers are increasingly investigating multiple access techniques for aerial networks.
Multiple access techniques can be categorized into OMA and NOMA schemes. In what follows, we summarize the most common OMA-based aBS schemes.


In OMA schemes, users are allocated orthogonal resources, e.g., time (TDMA), frequency (FDMA), codes (CDMA), or space (SDMA), as illustrated in Fig. \ref{Fig:OMA}. In \cite{Lyu2016}, the authors proposed a cyclical TDMA technique allowing an aBS, which follows a cyclical trajectory above gUEs, to maximize gUEs' minimal throughputs. To this end, they optimized the time allocated to each user during the aBS's flight. They showed  that there is a tradeoff between throughput and access delay, and further showed, through simulation results, that the proposed scheme is superior to static aBS placement. In \cite{Yin2019}, Yin \textit{et al.} investigated joint aBS placement, power and time duration allocations for TDMA and FDMA, aiming to maximize the system's common throughput, which was defined in \cite{Lyu2016}. They assumed that the aBS could be wirelessly charged by a gBS, and that the harvested power could be used for transmissions to gUEs. Numerical results showed that FDMA outperforms TDMA in terms of common throughput, due to shorter wireless charging duration. 
The work in \cite{Lyu2016} was extended in \cite{Wu2018} and \cite{Wu2017} to OFDMA-based aerial systems, where the authors characterized the common throughput-delay tradeoff. In \cite{Xue2019}, Xue \textit{et al.} proposed an energy-efficient OFDMA-based aerial network, where an aBS was deployed to serve wireless sensors. They proposed a solution to  the joint trajectory and OFDMA resource allocation problem. Numerical results demonstrated performance gains in terms of energy consumption compared to benchmark approaches.    

Unlike time- and frequency-based multiple access techniques, Ho \textit{et al.} proposed in \cite{Ho2010} and \cite{Ho2010_2} a prioritized frame selection-based CDMA scheme, where a set of active wireless sensors was divided into several groups, each of which communicated with the aBS using the CDMA technique. The objective was to maximize the number of transmit sensors, while minimizing the number of groups. Simulation results illustrated a small packet loss rate with  low mobility aBSs. Subsequently, the proposed framework was extended to Rician fading in \cite{Ho2011}.

Typically, SDMA consists of geographical partitioning of an area into multiple divisions, where each division is mapped to a resource. In wireless systems, SDMA leverages beamforming (or precoding) in order to create distinct communication beams aimed towards users, using the same resource. By doing so, more power can be directed to each user, hence, improving the quality of the communication link. To avoid inter-user interference, stringent precoding and power allocation design requirements are of great importance. It is worth noting that, due to the particular nature of aBSs, this task is more complex than that of gBSs. 

Several recent analyses have investigated SDMA in aerial communications. Jiang \textit{et al.} derived the optimal trajectory and heading of aBSs serving static gUEs in the context of an uplink SDMA scenario \cite{Jiang2012}. In \cite{Xie2014}, Xie \textit{et al.} proposed SDMA for L-band control communications between airborne stations (ASs) and a control center. By grouping ASs spatially, each group was allocated one radio resource, which was used randomly for access. By implementing SDMA within each group, ASs in the same group could communicate with the control center without causing any interference to others. The results obtained illustrated an improved {}{spectral efficiency} and reduced average delays. In \cite{Chen2018_2}, a novel time modulated array based SDMA uplink between gUEs and an aBS was proposed. Simulation and experimental results were presented to validate their proposal. 
The authors in \cite{Xiao2016} discussed optimal SDMA user grouping and precoding for mmWave uplink aBS-gUE communications.
In \cite{Ren2019}, Ren \textit{et al.} presented a machine learning (ML) hybrid precoding technique for massive MIMO aerial communications, where an aBS was equipped with massive MIMO operating in the mmWave frequency band. Aiming to maximize the EE, precoding was optimized using hybrid ML and error estimation optimization. Simulation results show that the proposed approach outperforms other benchmark techniques, such as zero-forcing (ZF) precoding, antenna selection, and adaptive cross-entropy. Finally, Tan \textit{et al.} derived in \cite{TAN2019} the exact expression of the achievable rate of a massive MIMO aBS operating in the mmWave frequency band under Ricean fading channel. Assuming CSI knowledge at the aBS, a statistical eigenmode-SDMA approach was proposed for the downlink transmission in a two-user MISO system. It has been found that the performance, in terms of achievable sum rate, saturates at high signal-to-interference-plus-noise ratio (SINR) regime. 
Table \ref{TableOMA} summarizes the main related work on OMA-based aerial networks.

\begin{table*}
\caption{Related Work on OMA-based Aerial Networks.}
\label{TableOMA}
\footnotesize
\begin{tabular}{|p{30pt}|p{90pt}|p{75pt}|p{100pt}|p{140pt}|}
\hline
{\textbf{Reference}} & \makecell{\textbf{System Model}} & \makecell{\textbf{Objective}} & \makecell{\textbf{Approach Followed}} & \makecell{\textbf{Findings}}  \\
\hline 
\hline
\makecell{\cite{Lyu2016}}	& \makecell{aBS + cyclical TDMA \\+ cyclical trajectory.} & \makecell{Max. minimum user \\throughput.} & \makecell{Optimization of allocated \\time to users.} & \makecell{There is a tradeoff between throughput \\and access delay. \\ aBS cyclical trajectory outperforms \\static placement.} \\ \hline
\makecell{\cite{Yin2019}} & \makecell{aBS + TDMA or FDMA \\+ energy harvesting \\for communication} & \makecell{Max. common \\throughput.} & \makecell{Joint aBS placement, power \\and time duration allocation.} & \makecell{FDMA outperforms TDMA, due to \\shorter wireless charging duration.} \\ \hline

\makecell{\cite{Wu2018,Wu2017}} & \makecell{aBS + OFDMA.} & \makecell{Max. minimum avg. \\ throughput of users.} & \makecell{Joint optimization of UAV \\trajectory and OFDMA \\resource allocation} 
& \makecell{Validation of the existence of \\a fundamental throughput-delay tradeoff.\\ Simulation results agree with the \\theoretical ones.}  \\ \hline

\makecell{\cite{Xue2019} } & \makecell{aBS + OFDMA.} & \makecell{Max. EE} &  \makecell{Joint optimization of UAV \\trajectory and OFDMA \\resource allocation.} & \makecell{Achieved EE is superior to \\that of benchmark approaches.} \\ \hline

\makecell{\cite{Ho2010,Ho2010_2}} & \makecell{aBS + uplink prioritized \\frame selection-based \\CDMA + nodes grouping.} & \makecell{Max. nbr. of tx. nodes\\ to aBS + Min. nbr. of \\groups.} &  \makecell{Fixed aBS trajectory \\+ different priority \\group divisions.} & \makecell{A small packet loss rate is achieved \\with low mobile aBS.} \\ \hline

\makecell{\cite{Ho2011}} & \makecell{aBS + uplink prioritized \\frame selection-based\\ CDMA + nodes grouping.} & \makecell{Min. packet error rate.} & \makecell{Number of priority groups \\and aBS altitude optimization,\\ considering Ricean channels.} & \makecell{Proposed protocol outperforms TDMA.}  \\ \hline

\makecell{\cite{Jiang2012}} & \makecell{Multi-user MISO aBS \\+ uplink SDMA.} & \makecell{Max. approx. ergodic\\ sum rate.} &  \makecell{aBS trajectory and heading \\optimization.} & \makecell{SDMA achieves better performances \\than TDMA.\\ Simplified optimization algorithms realize \\near-optimal results.} \\ \hline

\makecell{\cite{Xie2014}} & \makecell{ASs + uplink SDMA \\+ ASs grouping.} & \makecell{Max. SE.} & \makecell{Implementation of SDMA \\based collision free random \\access, with ZF beamforming.} & \makecell{Proposed access scheme improves \\the SE and reduces the average time delay.}  \\ \hline

\makecell{\cite{Chen2018_2}} & \makecell{Multi-user MISO aBS \\+ uplink time modulated \\SDMA.} & \makecell{Eval. per-user \\bit error rate.} & \makecell{Implementation of time \\modulated array with SDMA.} & \makecell{With a single RF chain, proposed scheme \\succeeds in separating the signals. \\Experiment validates simulation results.} \\ \hline

\makecell{\cite{Xiao2016}} & \makecell{Multi-user MISO aBS \\+ uplink SDMA \\+ users grouping.} & \makecell{Max. total achievable \\data rate.} & \makecell{Users grouping optimization\\and beamforming precoding.} & \makecell{Proposed mmWave-SDMA outperforms \\conventional systems in terms of total\\ achievable data rate.}  \\  \hline

\makecell{\cite{Ren2019}} & \makecell{Massive MIMO aBS \\+ mmWave \\+ downlink SDMA.} & \makecell{Max. EE.} & \makecell{Precoding optimization via \\hybrid ML/error estimation.} & \makecell{Proposed approach outperforms ZF \\precoding, antenna selection, and adaptive\\ cross-entropy benchmarks.}  \\ \hline

\makecell{\cite{TAN2019}} & \makecell{Multi-user MISO aBS \\+ mmWave + downlink \\stat. eigenmode-SDMA.} & \makecell{Eval. sum rate.} & \makecell{Beamforming precoding.} & \makecell{A closed-form expression of the achievable \\rate is derived.\\The achievable sum rate saturates for high \\SINRs (above 20 dB).}\\ \hline 
\end{tabular}
\end{table*}

\begin{figure}[t]
\centering
\includegraphics[width=0.95\linewidth]{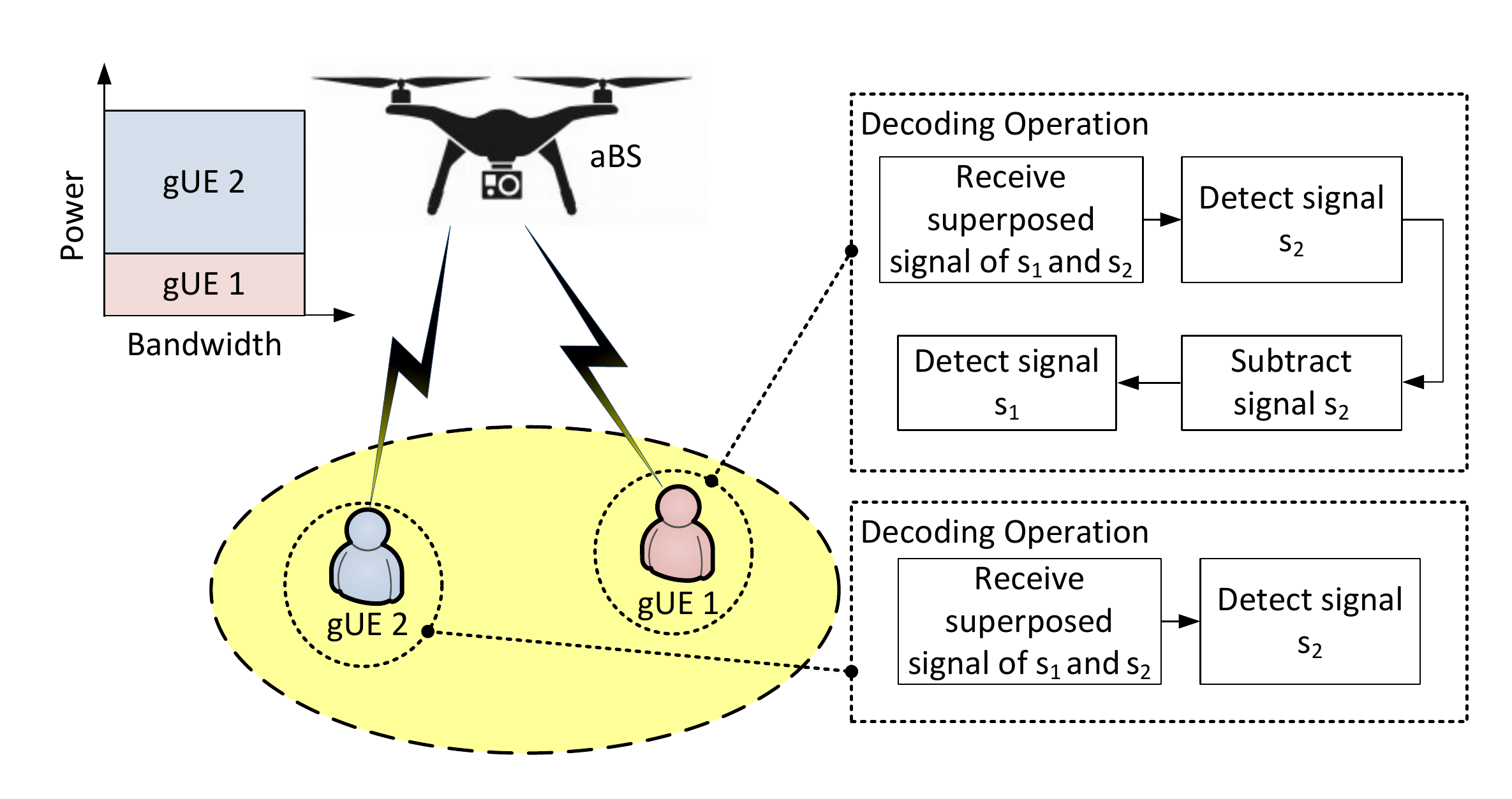}
\caption{Illustration of NOMA technique in a two-user aerial system.}
\label{Fig:NOMA}
\end{figure}

\section{Non-Orthogonal Multiple Access in Aerial Networks}
As discussed earlier, OMA techniques ensure orthogonality between concurrent communications in order to avoid or reduce interference among them. However, this type of design is limited by the number of users served, which, in turn, reduces the system's {}{spectral efficiency}. One approach to overcome this challenge is to utilize non-orthogonal transmissions. Recently, NOMA has emerged as a strong candidate and key enabling technology for 5G and beyond \cite{Ding2014}. Its key principle lies in allowing different users to share the same frequency/time resource blocks at the expense of inter-user interference. To combat this interference, different users are either assigned different codes, known as C-NOMA \cite{Ali2017,Wei2018}, or distinct power levels, referred to as P-NOMA \cite{Ding2014}.

In the  context of aerial networks, most research has focused on P-NOMA. 
\textcolor{black}{For the SISO two-user downlink scenario shown in Fig. \ref{Fig:NOMA}, P-NOMA works as follows. First, the aBS superposes users' signals $s_1$ and $s_2$ with power levels $P_1$ and $P_2$, such that the gUE with the weakest wireless link to the aBS gets a higher amount of power for its signal, in this case $P_2>P_1$. The superposed signal is then transmitted using the entire bandwidth. At gUE 1, SIC is used in order to decode signal $s_2$ at first, subtract it, then decode its own signal $s_1$, whereas gUE 2 decodes directly its signal $s_2$. It is to be noted that an uplink scenario using P-NOMA can be similarly described.}

Recent surveys presented an overview of the application of P-NOMA in future wireless networks (including UAVs) \cite{maraqa2019survey,Li2019}.
Meanwhile, more technical papers have appeared in the recent literature. The work of \cite{Liu2019_2} focused on NOMA within UAV networks, where joint trajectory design and power allocation is investigated, while an extension to a hybrid OMA/NOMA scheme for UAV-assisted vehicle-to-everything (V2X) networks was studied in \cite{Abbasi2020}. In \cite{Sohail2018}, Sohail \textit{et al.} addressed the sum rate maximization problem of a NOMA-based aBS downlink scheme. Optimal aBS placement was obtained at first, then power allocation coefficients were determined. In \cite{Liu2019,Sun2019}, a joint NOMA-aBS placement and power allocation problem was formulated and solved with a view to improving the system's sum rate and throughput, respectively. An extension to \cite{Sun2019} was reported in \cite{Nasir2019}, where aBS altitude, user scheduling, and transmit antenna beamwidth optimizations were considered. The authors of \cite{Nguyen2019_inrs} extended \cite{Liu2019} to joint NOMA power allocation, user pairing, and UAV placement to maximize the minimum sum rate of user pairs.
In \cite{Cui2019}, joint trajectory and power allocation for a NOMA-aBS serving ground users was investigated, aiming to maximize the minimum average rate, whereas \cite{Tang2020} improves the number of satisfied users with quality-of-expereince by optimizing the UAV placement, admission control, and NOMA power. For the uplink, authors in \cite{Seo2019} investigated NOMA random access (RA), where transmit power is controlled in order to allow the reception of one of two power levels at the aBS. Subsequently, they derived the maximum stable throughput as a function of the aBS altitude and beamwidth.

Although most reported results on NOMA have focused on single-antenna aBSs, there have been some recent results which considered NOMA-based MIMO aBSs. For example, in \cite{Hou2019}, Hou \textit{et al.} considered a NOMA MIMO aBS-gUEs downlink system, and further derived outage probability and ergodic data rate expressions for both LoS and NLoS channel models. In \cite{Rupa2019}, Rupasinghe \textit{et al.} leveraged MIMO to generate directional beams aimed towards users, given limited distance information feedback at the aBS. Users within the same beam were served using NOMA. The same authors extended this work in \cite{Rupa2019_2} to limited users angles information feedback at the aBS. Finally, cooperative NOMA has been proposed in \cite{Nguyen2018,Zhao2019_noma}. Nguyen \textit{et al.} exploited cooperative aBSs in \cite{Nguyen2018} to communicate with users through a virtually designed MIMO NOMA channel, whereas Zhao \textit{et al.} proposed in \cite{Zhao2019_noma} cooperation among aBSs and gBSs to serve ground users. Table \ref{TableNOMA} summarizes the main related work on NOMA-based aerial networks.


\begin{table*}
\caption{Related Work on NOMA-based Aerial Networks.}
\label{TableNOMA}
\footnotesize
\begin{tabular}{|p{30pt}|p{90pt}|p{75pt}|p{100pt}|p{140pt}|}
\hline
{\textbf{Reference}} & \makecell{\textbf{System Model}} & \makecell{\textbf{Objective}} & \makecell{\textbf{Approach Followed}} & \makecell{\textbf{Findings}}  \\
\hline 
\hline
\makecell{\cite{Liu2019_2}} & \makecell{Multi-user SISO aBS \\+ downlink NOMA.} & \makecell{Max. minimum \\user rate.} & \makecell{Joint power allocation and \\ trajectory design with ML.} & \makecell{NOMA outperforms OMA benchmark \\scheme.}  \\ \hline

\makecell{\cite{Abbasi2020}} & \makecell{Multi-user SISO aBS \\+ downlink hybrid \\OMA/NOMA in V2X.} & \makecell{Max. sum rate.\\Max. minimum rate.} & \makecell{Joint power allocation and \\ trajectory design using \\alternating optimization (AO).} & \makecell{NOMA has always a better or equal \\sum-rate than
OMA at high SNR regime.\\
Proposed dynamic OMA/NOMA scheme \\achieves less
decoding complexity than\\NOMA with a negligible sum-rate \\compromise.
}  \\ \hline

\makecell{\cite{Sohail2018} } & \makecell{Multi-user SISO aBS \\ + downlink NOMA.} & \makecell{Max. sum rate.} &  \makecell{Sequential aBS placement \\and power allocation.} & \makecell{NOMA is superior to OMA in terms of \\sum rate and EE.\\ Lower carrier frequency improves NOMA's \\performance gain.} \\ \hline

\makecell{\cite{Liu2019}} & \makecell{Multi-user SISO aBS \\ + downlink NOMA.} & \makecell{Max. sum rate.} &  \makecell{Sequential aBS placement \\optimization and power \\allocation.} & \makecell{Sequential optimization achieves \\better results than benchmark schemes.} \\ \hline

\makecell{\cite{Sun2019}} & \makecell{Multi-user SISO aBS \\ + downlink cyclical \\NOMA.} & \makecell{Max. minimum \\ throughput.} &  \makecell{AO of users scheduling \\and aBS trajectory.} & \makecell{Cyclical NOMA outperforms cyclical \\TDMA in terms of minimum throughput, \\average access delay, and flying range \\for the same minimum throughput.} \\ \hline

\makecell{\cite{Nasir2019}} & \makecell{Multi-user SISO aBS \\ + downlink NOMA.} & \makecell{Max. minimum rate.} & \makecell{Joint optimization of aBS \\altitude, transmit antenna \\beamwidth, bandwidth, \\and power, using \\a path-following algorithm.} & \makecell{NOMA is superior to OMA scheme. \\NOMA achieves rates close to \\dirty-paper-coding (DPC).}  \\ \hline

\makecell{\cite{Nguyen2019_inrs}} & \makecell{Multi-user SISO aBS \\ + downlink NOMA \\+ users pairing.} & \makecell{Max. sum rate.} & \makecell{Joint optimization of aBS \\altitude, user pairing \\and power using heuristic \\pairing based on the min. \\sum of squared distance \\ criteria.} & \makecell{Proposed user pairing scheme achieves \\near-optimal performance.}  \\ \hline

\makecell{\cite{Cui2019}} & \makecell{Multi-user SISO aBS \\ + downlink NOMA.} & \makecell{Max. minimum \\avg. rate.} &  \makecell{Joint trajectory and power \\allocation using penalty \\dual-decomposition technique.} & \makecell{Proposed approach outperforms\\ OMA and NOMA benchmarks.} \\ \hline

\makecell{\cite{Tang2020}} & \makecell{Multi-user SISO aBS \\ + downlink NOMA.} & \makecell{Max. number of \\satisfied users with \\quality-of-experience.} &  \makecell{Joint aBS placement, power\\ control, and admission control \\ based on the penalty
function \\method and successive convex \\approximation.} & \makecell{Proposed approach converges to \\a near-optimal solution within three \\iterations in polynomial time.} \\ \hline

\makecell{\cite{Seo2019}} & \makecell{Multi-user SISO aBS \\ + uplink NOMA RA.} & \makecell{Eval. throughput.} &  \makecell{Transmit power optimization.} & \makecell{Maximum stable throughput is a function \\of aBS altitude and beamwidth.\\ Proposed algorithm achieves throughput \\stabilization for dynamic traffic.} \\ \hline

\makecell{\cite{Hou2019}} & \makecell{Multi-user MIMO aBS \\ + downlink cluster-based \\NOMA.} & \makecell{Eval. ergodic rate \\and outage \\probability.} & \makecell{Stochastic geometry analysis.} & \makecell{Exact closed-form expressions for \\ergodic rate and outage probability are \\derived in both LoS and NLoS scenarios.\\ Performances are enhanced with LoS.}  \\ \hline

\makecell{\cite{Rupa2019}} & \makecell{Multi-user mmWave \\MIMO aBS + downlink \\ NOMA + partial CSI.} & \makecell{Max. achievable sum \\rate
and eval. \\outage probability.} & \makecell{aBS altitude optimization \\and directional beaming \\using beam scanning.} & \makecell{Distance feedback is an efficient alternative \\to full CSI feedback. \\ NOMA outperforms OMA benchmark.\\ Achievable sum rates depend on NOMA \\users selection.} \\ \hline

\makecell{\cite{Rupa2019_2}} & \makecell{Multi-user mmWave \\MIMO aBS + downlink \\ NOMA + partial CSI.} & \makecell{Eval. achievable sum \\rate
and outage \\probability.} & \makecell{Stochastic geometry and order \\statistics. \\Adoption of different user \\ordering strategies (Fejer \\kernel, and angle based \\ordering).} & \makecell{
It is important to identify under which \\feedback scheme (angles or distances)
\\users become more distinguishable. \\ 
Proposed users ordering strategies \\outperform classical distance ordering.} \\ \hline 

\makecell{\cite{Nguyen2018}} & \makecell{gBS + multi-user coop. \\virtual MIMO aBSs \\+ downlink NOMA.} & \makecell{Max. achievable sum\\ rate.} & \makecell{Joint optimization of radio \\resource allocation, NOMA \\decoding order, and aBSs \\locations, based on \\DC programming.} & \makecell{Proposed solution is superior to \\conventional approaches without coop. \\NOMA or aBSs locations optimization.}  \\ \hline

\makecell{\cite{Zhao2019_noma}} & \makecell{Multi-user gBS + aBS \\downlink NOMA.} & \makecell{Max. sum rate.} & \makecell{Alternate user scheduling \\and UAV trajectory \\optimization + NOMA\\ precoding.} & \makecell{aBSs should fly close to its served \\users and avoid gBSs served users \\to guarantee improved performance.}  \\ \hline
\end{tabular}
\end{table*}

\section{Rate-Splitting Multiple Access}
In the previous section, we discussed the main OMA and NOMA techniques along with their applications to aerial networks. 
Although NOMA enables simultaneous transmissions to a large number of users with correlated and non-orthogonal channels, i.e., an overloaded scenario, it may not be optimal for single-antenna multi-user systems. Indeed, if {}{a number of users smaller than the number of transmit antennas is scheduled}, NOMA would perform worse than SDMA due to unnecessary data combining in a wireless environment with only few users and light or moderate interference, i.e., an underloaded scenario. By contrast, SDMA performance would suffer in an overloaded scenario, where the number of {}{scheduled} users exceeds the number of {}{transmit antennas \cite{Mao2018}}. Consequently, a generalized configuration is needed in order to optimize the utilization of resources for any load scenario. This is what motivated the proposal of RSMA as a generalizing scheme of NOMA and SDMA.

\subsection{Background}
{}{The term RSMA has been initially proposed by Rimoldi \textit{et al.} in \cite{Rimoldi1996}, where, in order to achieve a general point in the capacity region of a Gaussian multiple access channel, a proper \textit{code} must be constructed for each user. This idea has been extended to discrete memoryless channels \cite{Grant2001}, and distributed rate-splitting \cite{Cao2007}. Nevertheless, most works were limited to a theoretical framework; yet, with the advent of multiple-antenna technology and low-complex SIC technique, RSMA has been living a new start \cite{Clerckx2016}. Indeed, it} is expected to achieve promising performance gains in both overloaded and underloaded scenarios \textcolor{black}{\cite{Clerckx2016,Mao2018}}. 


Similar to NOMA, RSMA relies on the implementation of a linear precoder at the transmitter and a SIC at the receiver. 
\textcolor{black}{In downlink RSMA, the} process begins with user messages \textcolor{black}{($M_1$ and $M_2$)} being divided into common and private parts \textcolor{black}{($M_i^c$ and $M_i^p$ with $i$ the index of the user)} at the transmitter \textcolor{black}{(e.g., aBS)}. The common parts of all users are combined together and encoded into a single common stream \textcolor{black}{($s_{12}$ for two users)}, where the private parts are encoded into distinct private streams \textcolor{black}{($s_1$ and $s_2$ for two users)}. Finally, the resulting streams are superimposed into one signal and sent over a MIMO channel. At each user, the common stream is first decoded and the user's own data is recovered. At the receiver, the interference from the common stream is removed using the SIC. This is followed by decoding the private parts of the user's message, while treating private parts of other user messages as noise. 

\textcolor{black}{In uplink RSMA, the process is slightly different. Specifically, each user splits its intended message ($M_i$) into common and private parts ($M_i^c$ and $M_i^p$) that are then encoded and linearly precoded before transmission. Since the receiver receives a combined signal of all common and private parts of all users' messages, it proceeds using SIC to decode the signals one by one, while treating the remaining signals as noise. It is worth noting that the decoding order at the receiver is crucial in determining its performance. The downlink and uplink RSMA mechanisms are illustrated in Fig. \ref{Fig:model} below.\footnote{\textcolor{black}{In Fig. \ref{Fig:model}b, an example of a decoding order is presented.}}}


\textcolor{black}{The capability of RSMA to split the message into a common part and a private part provides the flexibility to partially decode interference and partially treat interference as noise. This enables a soft bridging of two extremes, namely fully decoding interference (as in NOMA), and treating all interference as noise (as in SDMA), hence providing room for data rates and QoS improvements, besides complexity reduction. Moreover, RSMA mechanism can be useful for several applications, e.g., coded caching and multicast systems \cite{Fadlallah2017,Ding2018,Mao2018_2,Chen2020}. Focusing on the coded caching example, a media content can be subpacketized and packets are adequately cached in end-devices. Then, each RSMA-based coded content delivery would occupy a reduced number of resources, e.g., subcarriers, time slots, etc. Finally, cached data can be exploited for interference cancellation \cite{Xiang2018,Maddah2019}, which ultimately renders RSMA more spectrally- and energy-efficient.}


\begin{figure*}[t]
\centering
\includegraphics[width=0.999\linewidth]{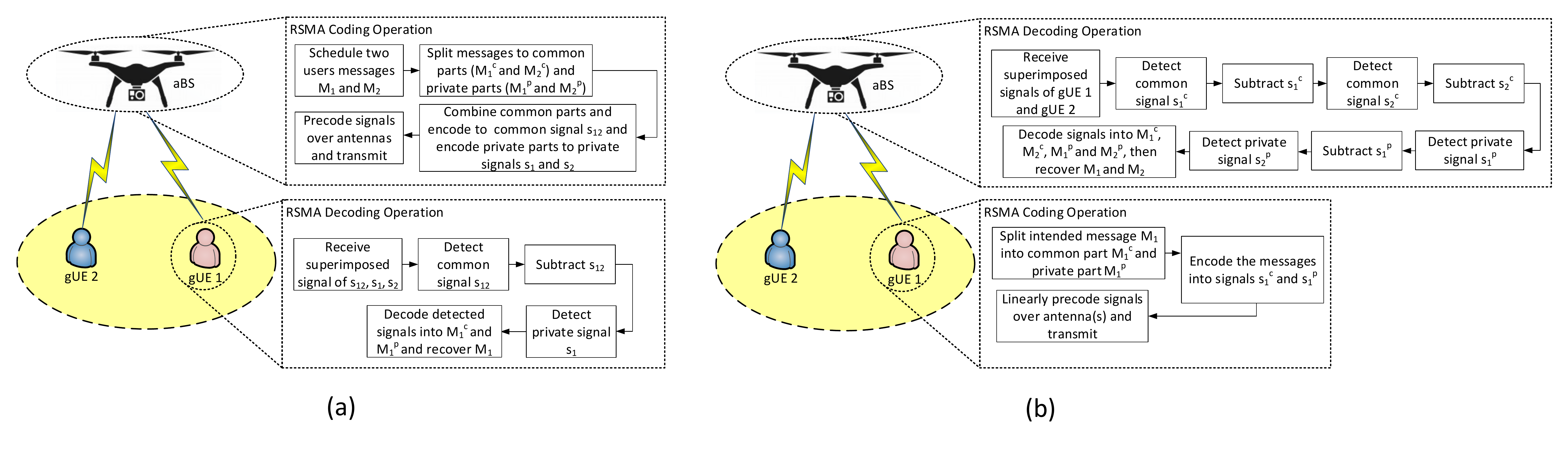}
\caption{\textcolor{black}{System model: (a) downlink, (b) uplink.}}
\label{Fig:model}
\end{figure*}

RSMA has attracted the attention of the research community recently, where several issues, e.g., rate-splitting, power allocation or precoding strategy for common and private streams, decoding order policy, and the designed order of common streams, have been extensively investigated. 
{}{In \cite{Hao2015}, Hao \textit{et al.} proposed a simple  encoding scheme for RSMA using the conventional ZF beamforming. They also evaluated the sum-rate of a two-user broadcast channel with limited channel state information feedback to the transmitter (CSIT). Joudeh \textit{et al.} investigated in \cite{Joudeh2016} the ergodic sum rate maximization of a multi-user MISO system using RS and under imperfect CSIT. They showed that RS is efficient even with a partial CSIT and outperforms conventional transmission schemes with no rate-splitting. This work has been extended in \cite{Joudeh2016_2} to achieving max-min fairness amongst users, by analyzing the available degrees of freedom. In \cite{Joudeh2017}, the authors proposed a hybrid precoding for RSMA messages, aiming to achieve max-min fairness among multiple co-channel multicast groups. The superiority of their approach was demonstrated through a degrees of freedom analysis and simulation. {The degrees of freedom region of the multi-user MISO broadcast channel with partial CSIT has been investigated in \cite{Joudeh2019}, where the achievable region is realized using the RS approach and employing the Fourier-Motzkin elimination.}
In \cite{Dai2016}, RSMA was considered for massive MIMO with imperfect CSIT. In particular, the authors proposed a {hierarchical rate-splitting} (HRS) framework, where two different types of common messages were defined, i.e, ones that could be decoded by all users, and ones that could only be decoded by a subset of users.
In \cite{Papaz2017}, Papazafeiropoulos \textit{et al.} evaluated the sum rate of RSMA in massive MIMO systems, in the presence of various impairments such as phase distortion, thermal noises, and the availability of perfect/imperfect CSIT.} 
{Recently, Su \textit{et al.} proposed a novel RSMA scheme where interference alignment (IA) based rate splitting (RS) is leveraged to mitigate interference in a K-user MIMO interference channel \cite{Su2020}. Additionally, sum rate, outage probability, and symbol error probability (SER) expressions were derived and evaluated under imperfect CSIT. It was concluded that the combination of IA and RS
provides robust and reliable multiple access transmissions.}
In \cite{Mao2018}, a robust analytical framework was presented, where the performance of RSMA in terrestrial MU-MISO broadcast channels was evaluated in terms of weighted sum-rate (WSR). Results showed that RSMA outperforms NOMA and SDMA in a wide range of network loads and user deployment scenarios, as well as in lower complexity receiver designs, compared to NOMA. 
In \cite{Mao2018_3} and \cite{Mao2018_2}, Mao \textit{et al.} demonstrated SE and EE improvements of RSMA over NOMA and SDMA in unicast and multicast scenarios. {}{In \cite{zhou2020ratesplitting}, Zhou \textit{et al.} addressed the SE and EE tradeoff by investigating the joint SE and EE maximization problem for a RSMA-based MISO broadcast channel. The corresponding results demonstrate the efficiency of their approach compared to non-RS strategies, in terms of SE, EE and their tradeoff.} 
{}{In \cite{Clerckx2019}, Clerckx \textit{et al.} have analytically shown the flexibility of RSMA in generalizing, and subsuming as special cases all of SDMA, OMA, NOMA and multicasting, in the context of a two-user multi-antenna broadcast channel.}
{Whereas, the authors of \cite{salem2019rate} studied RS under phase-shift-keying input codebook for a multi-antenna multi-user system, where an interference exploitation approach aiming to improve the sum-rate performance is proposed. In \cite{mao2019dirty}, the authors combined RS and dirty-paper-coding in order to enlarge the achievable rate region for a multi-user MISO broadcast channel experiencing imperfect CSIT.} {In \cite{dizdar2020ratesplitting2}, Dzidar \textit{et al.} designed the first RSMA in the physical layer, accounting for adaptive modulation, coding, message splitting, and SIC decoding. Through link-layer evaluation, they illustrated the robustness and throughput superiority of RSMA over SDMA and NOMA, for both unconstrained and constrained QoS requirements and with imperfect CSIT.}
{Mao \textit{et al.} investigated in \cite{mao2019maxmin} the joint optimization of precoding, message splitting, time allocation and relaying scheduling, aiming to maximize the minimum rate among cooperative users in a MISO broadcast channel. Their results showed the superiority of their proposed approach, compared to SDMA and non-RS schemes.}
{}{The authors in \cite{Papaz2018} proposed RS to tackle the saturation problem in MIMO cooperative systems. They demonstrated the robustness of RS for imperfect CSIT, which allowed to increase the self-interference range for half-duplex communications.}
In \cite{Yu2019}, a simplified RSMA technique for cloud radio access networks (C-RAN) was presented, where the number of created common messages increased linearly in accordance with the number of users, rather than exponentially. Through simulation results, they showed the superiority of their technique compared to NOMA and SDMA. 
{}{In a multi-user multi-antenna simultaneous wireless information and power transfer (SWIPT) context, the authors of \cite{Mao2019} investigated the application of RS for MISO SWIPT broadcast channels. They showed that by optimizing the RS precoders, their approach is able to outperform both NOMA and SDMA in terms of WSR in several deployment scenarios.}
{In \cite{mao2020ratesplitting}, the authors introduced RSMA to massive IoT networks. In order to achieve the best ergodic sum rate (ESR) performance, they proposed power portioning RSMA (PP-RSMA) and time-portioning RSMA (TP-RSMA). The former serves cellular and IoT user groups over orthogonal time slots, while the latter serves the two groups at the same time and non-orthogonally. It has been shown that PP-RSMA is superior to TP-RSMA and other baseline schemes in terms of ESR, and that it is robust to imperfect CSIT.}
{Xu \textit{et al.} proposed to integrate RSMA into a multi-antenna gBS capable of both serving ground users and detecting targets in specific azimuth angles in \cite{xu2020ratesplitting}. Yin \textit{et al.} advocated in \cite{yin2020ratesplitting} the use of RSMA for multibeam satellite communications. Their aim was to achieve max-min fairness of a multibeam multicast satellite system with imperfect CSIT. Through simulations, they demonstrated the superiority of RS compared to conventional non-RS approaches for satellite communications.}
However, in the context of aerial networks, there has been only few reported result on RSMA \cite{Ahmad2019,Rahmati2019}. In \cite{Ahmad2019}, RSMA-based C-RAN is realized using several gBSs and aBSs, whereas, the authors of \cite{Rahmati2019} reported the {}{energy efficiency} of an RSMA-based gBS towards aerial users. Finally, \cite{Yang2019_1,Yang2019_2} investigated the downlink and uplink SISO RSMA parameters optimization, supporting that their results would serve as a benchmark for the MIMO system. A summary of RSMA related work is presented in Table \ref{TableII}.

\begin{table*}
\caption{Related Work on RSMA-based Systems.}
\label{TableII}
\footnotesize
\begin{tabular}{|p{30pt}|p{90pt}|p{75pt}|p{100pt}|p{140pt}|}
\hline
{\textbf{Reference}} & \makecell{\textbf{System Model}} & \makecell{\textbf{Objective}} & \makecell{\textbf{Approach Followed}} & \makecell{\textbf{Findings}}  \\
\hline 
\hline
\makecell{\cite{Clerckx2016}\\ and \cite{Mao2018}}	& \makecell{Point-to-point MIMO. \\ Multi-user MISO\\ broadcast.} & \makecell{Def. users rate region. \\ Max. WSR.} & \makecell{Joint rate and beamforming\\ optimization using \\WMMSE-based AO.} & \makecell{RSMA design bridges the gap between \\ NOMA and SDMA. \\ RSMA outperforms NOMA and SDMA.} \\ \hline
\makecell{\cite{Mao2018_3}\\ and \cite{Mao2018_2}} & \makecell{Multi-user MISO \\unicast and multicast.} & \makecell{Max. SE (i.e., WSR). \\Max. EE.} & \makecell{Joint rate and beamforming\\ optimization using: \\WMMSE-based AO (WSR).\\ Successive Complex \\ Approximation (SCA)-based \\ algorithm (EE).} & \makecell{RSMA outperforms NOMA, SDMA,\\ and an OMA benchmark.} \\ \hline
\makecell{\cite{Hao2015}} & \makecell{Two-user MISO broadcast \\ + quantized CSIT.} & Max. ergodic sum rate. & \makecell{ZF beamforming  \\ of private streams \\+ space/time design \\of common streams.} & \makecell{A higher number of feedback bits leads \\ to a SNR/rate offset of the sum-rate. \\ Proposed scheme outperforms TDMA \\ and conventional ZF beamforming with \\ quantized CSIT.}  \\ \hline
\makecell{\cite{Joudeh2016} } & \makecell{Multi-user MISO  \\broadcast + partial CSIT.} & \makecell{Max. ergodic sum rate} &  \makecell{Joint rate and beamforming\\ optimization using: \\Sample Avg./ Conservatively \\ approximated WMMSE-based \\ AO.} & \makecell{RS allows relaxed CSIT quality. \\ RS enlarges the achievable ergodic \\ rate region.} \\ \hline
\makecell{\cite{Joudeh2016_2}} & \makecell{Multi-user MISO  \\broadcast + bounded CSIT \\errors.} & \makecell{Max-min fairness} &  \makecell{Joint rate and beamforming\\ optimization using: \\cutting-set method + \\ conservatively approximated \\ WMMSE)} & \makecell{A new approach to RS design.\\ RS eliminates the feasibility issue \\ arising in Non-RS designs.} \\ \hline
\makecell{\cite{Joudeh2017}} & \makecell{Overloaded multicast \\ groups.} & \makecell{Max-min. fairness} & \makecell{Joint rate and beamforming \\optimization using \\WMMSE-based AO.} & \makecell{RS outperforms both the proposed \\degraded and classical beamforming \\strategies.}  \\ \hline
\makecell{\cite{Joudeh2019}} & \makecell{Multi-user MISO\\ broadcast + partial CSIT.} & \makecell{Degrees of freedom \\ region \\characterization.} &  \makecell{Auxiliary variable elimination \\ (Fourier–Motzkin) approach.} & \makecell{Tuning a single power variable \\ and assigning a common degree of freedom \\ is enough to achieve all the \\ degrees of freedom region.} \\ \hline
\makecell{\cite{Dai2016}} & \makecell{Multi-user massive MIMO \\broadcast + imperfect \\CSIT.} & \makecell{Max. asymptotic \\ sum rate.} & \makecell{Proposed HRS design \\+ ZF beamforming \\+ power allocation.} & \makecell{HRS and RS are suitable for \\ massive MIMO.\\ HRS outperforms all other techniques\\ in presence of CSIT errors.}  \\ \hline
\makecell{\cite{Papaz2017}} & \makecell{Multi-user MISO \\broadcast \\+ hardware impairment \\+ perfect/imperfect CSIT.} & \makecell{Max. sum rate.} & \makecell{Distinct common and private \\ messages regularized-ZF \\ precoding + power allocation.} & \makecell{RS outperforms non-RS strategies, when \\ either additive hardware impairments are\\ present or absent, in conjunction with phase \\ and amplified thermal noises.\\ Increasing the number of users reduces the \\performance gain over non-RS techniques.} \\ \hline 
\makecell{{\cite{Su2020}}} & \makecell{{Multi-user MIMO} \\ {broadcast + imperfect} \\ { CSIT.}} & \makecell{{Eval. sum rate, outage} \\ {probability, and SER.}} & \makecell{{Stochastic geometry} \\ {and asymptotic analysis} \\ {of the proposed IA + RS} \\ {transmission scheme.}} & \makecell{{Relation between
RS and IA is identified.}\\ {Sum rate of RS is improved using IA} \\ {on common and private signals.}\\ {IA improves the outage probability}\\ {significantly.} \\ {SER is enhanced using a novel RS-based} \\ {adaptive modulation scheme.}} \\ \hline 
\makecell{\cite{zhou2020ratesplitting}} & \makecell{Multi-user MISO \\broadcast.} & \makecell{Max. jointly \\SE and EE.} & \makecell{Weighted -sum/-power \\ method + SCA-based \\algorithm.} & \makecell{RS outperforms non-RS strategies \\in terms of SE, EE and tradeoff.}  \\ \hline
\makecell{\cite{Clerckx2019}} & \makecell{Two-user MISO broadcast.} & \makecell{Derive tractable \\ sum-rate expressions.} & \makecell{Consideration of different \\ SNR regimes \\+ approximations.} & \makecell{RSMA unifies SDMA, OMA, NOMA \\and multicasting in one single approach.}  \\ \hline
\makecell{\cite{salem2019rate}} & \makecell{Multi-user MISO \\broadcast + phase-shift \\keying \\ + perfect/imperfect CSIT.} & \makecell{Max. sum-rate.} & \makecell{Interference suppression ZF \\ precoding. \\  Constructive-interference\\ precoding.} & \makecell{RS + constructive-interference precoding \\ outperforms conventional RS + ZF \\ precoding and non-RS approaches.}  \\ \hline
\makecell{\cite{mao2019dirty}} & \makecell{Multi-user MISO \\broadcast \\ + dirty-paper-coding \\ + partial CSIT.} & \makecell{Max. weighted \\ ergodic sum-rate.} & \makecell{Joint rate and beamforming \\optimization using \\WMMSE-based AO.} & \makecell{RS + dirty-paper-coding achieves \\ larger rate regions than non-RS  \\dirty-paper-coding for imperfect \\CSIT channels.}  \\ \hline
\makecell{{\cite{dizdar2020ratesplitting2}}} & \makecell{{Multi-user MISO} \\ {broadcast + AMC}  \\ {+ imperfect CSIT.}} & \makecell{{Max. throughput.}} & \makecell{{Design of
transmitter} \\ {and receiver architectures}  \\ {(finite constellation modul.} \\ {schemes, polar codes,} \\  {and an
AMC algorithm)} \\ {+ RSMA optimization} \\ {using WMMSE-based AO.}} & \makecell{{RSMA outperforms both SDMA} \\ {and NOMA in terms of throughput,} \\ {with and without QoS constraints.}}  \\ \hline
\end{tabular}
\end{table*}

 RSMA is more robust
and achieves significantly higher throughput than SDMA and
NOMA with and without QoS constraints in the considered
imperfect CSIT settings.

\begin{table*}
\label{TableIII}
\footnotesize
\begin{tabular}{|p{30pt}|p{90pt}|p{75pt}|p{100pt}|p{140pt}|}
\hline
{\textbf{Reference}} & \makecell{\textbf{System Model}} & \makecell{\textbf{Objective}} & \makecell{\textbf{Approach Followed}} & \makecell{\textbf{Findings}}  \\
\hline 
\hline
\makecell{\cite{mao2019maxmin}} & \makecell{Multi-user cooperative \\ MISO broadcast.} & \makecell{Max. min. sum-rate.} & \makecell{Self-organizing relaying \\ + SCA-based algorithm.} & \makecell{Proposed cooperative RS approach \\outperforms SDMA and non-RS schemes \\ for underloaded and overloaded networks, \\ and for different user deployment scenarios.}  \\ \hline
\makecell{\cite{Papaz2018}} & \makecell{Multi-pair massive \\ MIMO system.} & \makecell{Max. sum-rate.} & \makecell{Regularized ZF precoding.} & \makecell{RS is robust for both full-duplex \\and half-duplex relaying.\\ RS's robustness improves with a higher \\ number of relaying antennas  \\  or lower self-interference.\\ For a very high number of users, \\ RS's robustness degrades.}  \\ \hline
\makecell{\cite{Yu2019}} & \makecell{C-RAN of 1 baseband \\processing unit \\+ $N_R$ remote radio heads \\+ $N_U$ users.} & \makecell{Max. avg. min. user \\ rate.} & \makecell{WMMSE for RS, precoding \\ and fronthaul compression \\ optimization + hierarchical\\ users clustering.} & \makecell{Proposed RSMA
scheme outperforms \\ NOMA, SDMA, and conventional RSMA \\ schemes with one common stream or with \\ the same number of randomly selected \\ common streams.} \\ \hline
\makecell{\cite{Mao2019}}	& \makecell{Multi-user MISO SWIPT \\ broadcast.} & \makecell{Max. WSR.} & \makecell{Joint rate and precoding \\optimization using SCA \\+ WMMSE-based AO} & \makecell{Proposed RS-assisted approach achieves \\a better rate-energy tradeoff for SWIPT\\ downlink compared to SDMA and NOMA.} \\ \hline
\makecell{{\cite{mao2020ratesplitting}}}	& \makecell{{Multi-user MISO}  \\ {broadcast + two users} \\ {groups (cellular} \\ {+ massive IoT).}} & \makecell{{Max. ESR,}\\ {Characterize degrees} \\ {of freedom.}} & \makecell{{RSMA optimization using} \\ {sample
average approximation} \\ {+ WMMSE-based AO.}} & \makecell{{PP-RSMA is a powerful and general} \\ {transmission strategy for massive} \\ {overloaded cellular IoT networks.}\\
{PP-RSMA outperforms TP-RSMA and} \\ {other baseline multiple access schemes,} \\ {in terms of ESR.}} \\ \hline
\makecell{\cite{xu2020ratesplitting}} & \makecell{Multi-antenna radar \\communication system  \\ (downlink communication  \\ to users \\+ target detection).} & \makecell{Max. WSR + Approx. \\ beam-pattern to \\desired radar \\beam-pattern.} & \makecell{Joint rate and precoding \\optimization using an \\alternating direction method \\of multipliers.} & \makecell{RSMA achieves a better  \\WSR-beam-pattern  approximation \\tradeoff than SDMA.} \\ \hline
\makecell{\cite{yin2020ratesplitting}} & \makecell{Multibeam multicast \\ satellite system \\+ imperfect CSIT.} & \makecell{Max-min fairness.} & \makecell{Joint rate and beamforming \\optimization using modified\\ WMMSE-based AO.} & \makecell{Proposed RS-based approach efficiently \\ manages inter-beam interference and \\ support imperfect CSIT, per-feed \\ constraints and overloaded regime. \\ It outperforms non-RS schemes.} \\ \hline
\makecell{\cite{Ahmad2019}} & \makecell{C-RAN of gBSs and aBSs} & \makecell{Max. WSR.} & \makecell{Joint rate and beamforming \\optimization using the \\ convex-concave procedure.} & \makecell{Proposed method outperforms benchmark\\ approaches.\\ Preferred RS allocation assigns common \\messages to aBSs and private messages \\ to gBSs.}  \\ \hline

\makecell{\cite{Rahmati2019}} & \makecell{mmWave gBS \\ + aerial users.} & \makecell{Max. EE.} & \makecell{Joint rate and beamforming \\optimization using \\an SCA-based algorithm.} & \makecell{EE varies non-monotonically with aerial \\users' altitude. \\ RSMA outperforms NOMA.}  \\ \hline

\makecell{\cite{Yang2019_1}} & \makecell{Multi-user SISO downlink.} & \makecell{Max. sum rate.} & \makecell{Joint rate allocation and\\ power control optimization \\using a one-dimensional \\search algorithm.} & \makecell{\textcolor{black}{Given two users, RSMA and NOMA} \\ \textcolor{black}{realize the same exact capacity-achieving} \\ \textcolor{black}{sum rate.}\\
\textcolor{black}{For multiple users,} RSMA realizes up to \\15.6\% and 21.5\% \\sum rate gains compared\\ to NOMA and OFDMA, \textcolor{black}{respectively, given} \\ \textcolor{black}{one SIC layer only for RSMA and NOMA.}}  \\ \hline

\makecell{\cite{Yang2019_2}} & \makecell{Multi-user SISO uplink.} & \makecell{Max. sum rate.} & \makecell{Users transmit power\\and gBS decoding order \\optimization.} & \makecell{RSMA achieves up to 10.0\%, 22.2\%, \\and 83.7\% gains in terms of sum rate \\compared to NOMA, FDMA, and TDMA.}  \\ \hline

\end{tabular}
\end{table*}

Despite the extensive literature on RSMA in wireless networks, to the best of our knowledge, its use in aBSs is still in its infancy and has not yet been well investigated. 
Therefore, we provide here a look into RSMA-based aerial systems. In particular, we analyze and evaluate the WSR performance of a two-user system served using an RSMA-based aBS. Furthermore, a discussion of associated challenges and future research directions is provided.

\subsection{RSMA-Based Access in Aerial Networks}
\textcolor{black}{Here, we assume a simple RSMA-based downlink access from an aBS to two ground users, where the problems of RSMA parameters optimization and gBS placement are separately investigated. It should be noted that the joint optimization of RSMA parameters and UAV  placement achieves the full potential of RSMA in aerial networks. However, the investigation of this problem is beyond the scope of this survey.}

\subsubsection{System Model}
RSMA is a new paradigm that is expected to support massive connectivity in aerial networks. Fig. \ref{Fig:model} illustrates a simple scenario of RSMA-based aBS-gUEs downlink.
In our setup, we consider a $N$-antenna aBS serving two users, gUE1 and gUE2. The message of user $u$, denoted $M_u$, is divided into common parts $M_u^c$ and a private part $M_u^p$, {$\forall u \in \{1,2\}$}. The common parts are combined and encoded together to form one common stream, denoted by $s_{12}$. Finally, $s_u$ is the private stream of the encoded private message $M_u^p$, $\forall u \in \{1,2\}$. 
The resulting streams are then assigned precoding weights and superposed for transmission, assuming that the aBS has perfect channel knowledge. Hence, the received signal at user $u$ can be expressed by
\begin{equation}
    \label{eq:rx}
    y_u=\mathbf{h}_u^H \mathbf{P} \mathbf{s} + n_u,\quad \forall u \in \{1,2\}
\end{equation}
where $n_u$ is the complex additive white Gaussian noise (AWGN) with zero-mean and variance \textcolor{black}{$\sigma^2$}, $\mathbf{s}=[s_{12}, s_1, s_2]^t$ is the transmitted signal, $\mathbf{P}=\left[ \mathbf{p}_{12},\mathbf{p}_{1},\mathbf{p}_{2} \right] \in \mathbb{C}^{N \times 3}$ is the precoding matrix, and $\mathbf{h}_u=[h_u^1, \ldots, h_u^{N}]^t \in \mathbb{C}^{N \times 1}$ is the air-to-ground channel from the aBS to user $u$, and assumed to be perfectly known at the transmitter. The channel coefficient $h_u^i$, $\forall i=1,\ldots,N$, follows a Rician distribution with parameter $K_u$, given by 
\begin{equation}
    \label{eq:Ricechannel}
    h_u^{i}={d_u^{-\frac{\alpha}{2}}} g_u^i,
    \quad \forall i=1,\ldots,N, 
\end{equation}
where 
\begin{equation}
\alpha=a P_{\rm{LoS}}(\theta_u)+b
\end{equation}
is the path-loss factor \cite{Azari2018}, $\theta_u$ is the elevation angle between the aBS and user $u$, $g_u^i$ is defined as in (\ref{eq:smallscale}), $d_u=|\mathbf{w}-\mathbf{w}_u|$ is the distance between the aBS and user $u$, $\mathbf{w}=[x,y,z]$ and $\mathbf{w}_u=[x_u,y_u,0]$ are the aBS and user $u$ 3D locations, respectively, $\alpha$ is the path-loss factor, $\bar{h}_u^i$ is the deterministic LoS component with $|\bar{h}_u^i|=1$, and $\tilde{h}_u^i$ is the random NLoS component, modeled as a complex Gaussian distribution $\mathcal{CN}(0,1)$. Finally, $|.|$, $(.)^t$ and $(.)^H$ denote the Euclidian norm, transpose and conjugate transpose operators, respectively. {For the sake of clarity, we summarize in Table \ref{notationstable} the key notations.}

\begin{center}
\begin{table}[t]
\textcolor{black}{
\caption{\textcolor{black}{List of Key Notations.}}
\centering
\small
\begin{tabular}{|l|p{0.61\linewidth}|}
\hline \textbf{Symbol} & \textbf{Description}\\
\hline $\alpha$ & Path-loss exponent \\
\hline $\beta_{\rm LoS}$, $\beta_{\rm NLoS}$ & LoS and NLoS channel components\\
\hline $\gamma_u^{12}$, $\gamma_u^u$ & Received SINRs at gUE $u$ for the common and private streams \\
\hline $a,b,a_1,b_1$ & Channel modelling parameters\\
\hline $B$ & Transmission bandwidth\\
\hline $d_u$ & Distance between the aBS and gUE $u$ \\
\hline $g_u$ & Average aBS-gUE $u$ channel gain  \\
\hline $K$, $K_u$ & Rician K-factor and Rician K-factor for channel between aBS and gUE $u$\\
\hline $L$ & Number of potential aBS placement locations\\
\hline $M_u$, $M_u^c$, $M_u^p$ & gUE $u$'s message, common part message, and private part message\\
\hline $N$ & Number of antennas at the aBS\\
\hline $n_u$ & AWGN with zero mean and $\sigma^2$ variance at gUE $u$ \\
\hline $P_t$ & Maximal transmit power of the aBS\\
\hline $R_u^j$ & Achieved data rate at gUE $u$ for the $j^{th}$ stream transmission \\
\hline $R_{12}$ & Maximum allowed common stream data rate \\
\hline $R_u^{\rm com}$ & gUE $u$'s allocated portion of $R_{12}$ \\
\hline $R_u^{\rm tot}$ & gUE $u$'s total data rate (for common+private streams data rates) \\
\hline $\bar{R}$ & Weighted sum rate \\
\hline $R_u^{\rm th}$ & Minimum required data rate for successful decoding at gUE $u$\\
\hline $s_{12}$, $s_u$ & Common and private data streams (signals)\\
\hline $\mathbf{s}$ & Vector of transmitted streams \\
\hline $v_u$ & Traffic priority weight factor for gUE $u$ \\
\hline $\mathbf{h}_u=[h_u^1,\ldots,h_u^N]$ & Air-to-ground channel vector between aBS and gUE $u$ \\
\hline $\mathbf{P}=[\mathbf{p}_{12},\mathbf{p}_{1},\mathbf{p}_{2}]$ & aBS precoding matrix \\
\hline $\mathbf{r}=[R_1^{\rm com},R_2^{\rm com}]$ & aBS common data rate splitting strategy \\
\hline $\mathbf{w}$, $\mathbf{w}_u$ & 3D Cartesian locations of aBS and gUE $u$ \\
\hline
		\end{tabular}
		\label{notationstable} 
}
	\end{table}
\end{center}

\noindent
At user $u$, the signal {$s_{12}$} is decoded first, with the rest of the signal considered to be noise. Then, the latter is subtracted from the original signal using SIC. {}{Finally, $s_u$ is decoded, $\forall u \in \{1,2\}$ \cite{Mao2018}. The resulting Signal-to-Interference-plus-Noise Ratios (SINRs), denoted by $\gamma_u^{12}$ and $\gamma_u^{u}$, are written as
\begin{equation}
\label{eq:SINR12}
\gamma_u^{12}=\frac{|\mathbf{h}_u^H \mathbf{p}_{12}|^2}{|\mathbf{h}_u^H \mathbf{p}_1|^2+|\mathbf{h}_u^H \mathbf{p}_2|^2+\textcolor{black}{\sigma^2}},\quad \forall u \in \{1,2\},
\end{equation}
and
\begin{equation}
\label{eq:SINRu}
\gamma_u^{u}=\frac{|\mathbf{h}_u^H \mathbf{p}_{u}|^2}{|\mathbf{h}_u^H \mathbf{p}_{\bar{u}}|^2+\textcolor{black}{\sigma^2}},\quad \forall (u,\bar{u}) \in \{(1,2),(2,1)\},
\end{equation}
where the superscript designates the common or private decoded signal. Thus, the corresponding data rates are given by 
\begin{equation}
R_u^{j}=B\; \text{log}\left(1 + \gamma_u^j \right),    
\end{equation}
where $B$ is the bandwidth, $u \in \{1,2\}$, and $j \in \{12,1,2\}$. In order to decode the common stream $s_{12}$ at both users, the common rate shall not exceed $R_{12}=\text{min}(R^{12}_1,R^{12}_2)$. Rate boundaries for two-user rate splitting region is achieved if $R_{12}$ is adequately shared between the two users, i.e., $R_{12}=\sum_{u=1}^2 R_{u}^{\rm com}$, where $R_{u}^{\rm com}$ is user $u$'s portion of the common rate. Consequently, 
the total data rate of user $u$, denoted by $R_{u}^{\rm tot}$, can be expressed by \cite{Mao2018}
\begin{equation}
\label{eq:c4}
R_{u}^{\rm tot}=R_{u}^{\rm com}+R_u^u, \quad \forall u \in \{1,2\}.
\end{equation}
}

\subsubsection{Problem Formulation and Proposed Solution}

Deploying an aerial system brings several interesting challenges, including that of finding the optimal location of the aBS and its trajectory so that a desirable QoS can be achieved. Additional challenges may include how to determine the rate-splitting approach, precoding weights, number of SIC layers, and decoding order for accurate operations and improved performance. In this work, we attempt to briefly answer some of these questions. For clarity, we investigate a simple problem of WSR maximization by optimizing the aBS placement, rate-splitting, and precoding\footnote{{Although other performance metrics can be considered (e.g., coverage or delay), we focus here on the WSR metric due to its relevance in scarce-resource systems, and due to its established relation to other performance metrics.}}. 

In an aerial network, finding the optimal location of an aBS, given the statistical knowledge of channel conditions, is of paramount importance. In particular, based on the channel characteristics, one can determine the average weighted aBS location, which is evaluated as a function of the distances between the aBS and users, as well as the K-factors of the underlying fading channels. 
{}{
For any Rician fading channel $h_u^i$ ($u \in \{1,2\}$, $i=1,\ldots,N$), the average channel power is expressed by
\begin{equation}
g_u=\mathbb{E}(|h_u^i|^2)=\sqrt{\frac{\pi}{2}} d_u^{-\alpha/2} L_{\frac{1}{2}}(-K_u),    
\end{equation}
where $L_{\frac{1}{2}}$ is the Laguerre polynomial. As in \cite{Liu2019}, the aBS's best location can be obtained through the maximization of the weighted sum of average channel powers. This problem is similar to problem (12) in \cite{Liu2019}. Hence, it can be reformulated as the minimization of weighted sum of squared distances, i.e., $\min_{\mathbf{w}}$ $\sum_{u=1}^2 \lambda_u |\mathbf{w}-\mathbf{w}_u|^2$, where $\lambda_u=g_u/\left(g_1+g_2\right)$. Consequently, by following steps similar to Lemmas 1 and 2 in \cite{Liu2019}, the best statistical \textcolor{black}{and stationary} aBS location is given by
\begin{subequations}
    \label{eq:optCoord}
    \begin{equation}
    x_{\rm{s}}=\frac{\sum_{u=1}^2 \lambda_u x_u}{\lambda_{1}+\lambda_{2}},
    \end{equation}
    \begin{equation}
    y_{\rm{s}}=\frac{\sum_{u=1}^2 \lambda_u y_u}{\lambda_{1}+\lambda_{2}},
    \end{equation}
    \begin{equation}
    z_{\rm{s}}=z_{aBS},
    \end{equation}
\end{subequations}
where $z_{aBS}$ is the minimum flying altitude of the aBS\footnote{\textcolor{black}{To be noted that the best altitude is always the minimum tolerated one in an interference-free environment, i.e., where no other system using the same frequency band is present in the same area, since it allows to improve the channel's quality by reducing the distance to users.}}.}
However, this solution is sub-optimal, since it ignores the small-scale variations of the air-to-ground channels. Taking the latter into account, the problem becomes very complex and hard to solve, due to the dependency of the channel gain on the aBS location (according to (\ref{eq:smallscale})). 
To obtain a near-optimal solution, we opt for an \textit{iterative search} where several potential aBSs locations are evaluated by optimizing their associated RSMA parameters. Then, the location providing the best WSR performance is selected. For simplicity's sake, we expect that the best UAV location to be within the line segment linking the two users, at altitude $z_{aBS}$. We define $L$ equidistant points on this segment as potential aBS locations. It is obvious as $L \xrightarrow{} \infty$, the obtained solution becomes optimal. \textcolor{black}{The details of the iterative search will be presented later.} 

{Subsequently, the rate-splitting and precoding problem is formulated, as in \cite{Mao2018}. Given a known aBS location (among the defined potential locations), we optimize here the beamforming matrix $\mathbf{P}$ and the common rate vector $\mathbf{r}=[R_{1}^{\rm com},R_{2}^{\rm com}]$, aiming to maximize the WSR, defined as 
\begin{equation}
\bar{R}= v_1 R_{1}^{\rm tot}+v_2 R_2^{\rm tot},
\end{equation}
where $v_u$ ($u=1,2$) is the weight factor reflecting user $u$ traffic priority. The problem (P1) can be formulated as follows 
\begin{subequations}
	\begin{align}
	\small
	\max_{\mathbf{P}, \mathbf{r}} & \quad 
	{\bar{R}} \tag{P1} \\
	\label{c_P1}
	\text{s.t.}\quad & R_{1}^{\rm com}+R_2^{\rm com}\leq R_{12},  \nonumber \tag{P1.a} \\
	\label{c2} & R_{u}^{\rm tot}\geq R_{u}^{\rm{th}},\;\forall u \in \{1,2\} \tag{P1.b}\\
	\label{c3} & \text{tr}(\mathbf{P}\mathbf{P}^H)\leq P_t  \tag{P1.c}\\
	\label{c4} & \mathbf{r}\geq \mathbf{0}  \tag{P1.d} 
	\end{align}
\end{subequations}
where $R_{u}^{\rm{th}}$ is the minimum required data rate to ensure successful decoding. 
Problem (P1) is non-convex as demonstrated in \textcolor{black}{\cite{Joudeh2016,Christensen2008}. To solve it, we adopt the WMMSE-based AO approach that relies on transforming the problem to an augmented weighted MSE problem (AWMSE),} \textcolor{black}{detailed as follows} \textcolor{black}{\cite{Joudeh2016,Mao2018}}. 
\textcolor{black}{
gUE $u$ ($u \in \{1,2\}$) detects and estimates $s_{12}$ as $\hat{s}_{12}=f_u^{12}y_u$, where $f_u^{12}$ is the equalizer. After successfully decoding $s_{12}$ and subtracting it from the received signal, $s_u$ can be detected and estimated as
\setcounter{equation}{15}
\begin{equation}
\hat{s}_u=f_u^u (y_1-\textbf{h}_u^H \textbf{p}_{12}s_{12}).    
\end{equation}
We define the MSE of each stream as $\varepsilon_k=\mathbb{E}\{|s_k-\hat{s}_k|^2\}$, calculated as
\begin{equation}
\label{eq:equalizer}
\varepsilon_u^{12}=|f_u^{12}|^2 T_u^{12}- 2 \Re(f_u^{12}\textbf{h}_u^H \textbf{p}_{12})+1
\end{equation}
and
\begin{equation}
\label{eq:equalizer2}
\varepsilon_u^{u}=|f_u^{u}|^2 T_u^{u}- 2 \Re(f_u^{u}\textbf{h}_u^H \textbf{p}_{u})+1,
\end{equation}
where 
\begin{equation}
T_u^{12}=\sum_{i \in \{12,1,2 \}} |{\mathbf{h}}_u^H \textbf{p}_i|^2+\sigma^2   
\end{equation}
and 
\begin{equation}
T_u^u=T_u^{12}-|\textbf{h}_u^H \textbf{p}_{12}|^2+\sigma^2    
\end{equation}
are the received power at gUE $u$ to decode $s_{12}$ and $s_u$, respectively, and $\Re(\cdot)$ is the real part of a complex number. The optimal MMSE equalizers can then be given by \cite{Joudeh2016}
\begin{equation}
\label{eq:equalizeropt}
(f_u^{12})^{\rm MMSE}= \textbf{p}_{12}^H \textbf{h}_u (T_u^{12})^{-1}
\end{equation}
and
\begin{equation}
\label{eq:equalizeropt2}
(f_u^{u})^{\rm MMSE}= \textbf{p}_{u}^H \textbf{h}_u (T_u^{u})^{-1}.
\end{equation}
By substituting (\ref{eq:equalizeropt})--(\ref{eq:equalizeropt2}) into (\ref{eq:equalizer})--(\ref{eq:equalizer2}), the MMSEs can be written as
\begin{equation}
(\varepsilon_u^{12})^{\rm MMSE}=\min_{f_u^{12}} \varepsilon_u^{12}=(T_u^{12})^{-1} I_u^{12} 
\end{equation}
and
\begin{equation}
(\varepsilon_u^{u})^{\rm MMSE}=(T_u^{u})^{-1} I_u^{u},
\end{equation}
where 
\begin{equation}
 I_u^{12}=T_u^{12}-|\textbf{h}_u^H \textbf{p}_{12}|^2
\end{equation}
and 
\begin{equation}
I_u^{u}=T_u^{u}-|\textbf{h}_u^H \textbf{p}_{u}|^2    
\end{equation}
are the experienced interference when decoding signals $s_{12}$ and $s_u$, respectively. Thus, the SINRs can be expressed by
\begin{equation}
 \gamma_u^{12}=1/((\varepsilon_u^{12})^{\rm MMSE})-1 
\end{equation}
and 
\begin{equation}
 \gamma_u^{u}=1/((\varepsilon_u^{u})^{\rm MMSE})-1, 
\end{equation}
and the common and private data rates by
\begin{equation}
 R_u^{12}=-\log_2\left( (\varepsilon_u^{12})^{\rm MMSE}\right)   
\end{equation}
and 
\begin{equation}
 R_u^{u}=-\log_2\left( (\varepsilon_u^{u})^{\rm MMSE}\right),   
\end{equation}
respectively. Consequently, the AWMSEs can be expressed by
\begin{equation}
    \label{eq:awmse}
    \zeta_u^{12}= \mu_u^{12} \varepsilon_u^{12} -\log_2\left( \mu_u^{12} \right)
    \end{equation}
    and 
    \begin{equation}
    \label{eq:awmse2}
    \zeta_u^{u}= \mu_u^{u} \varepsilon_u^{u} -\log_2\left( \mu_u^{u} \right),
\end{equation}
where $\mu_u^{j}>0$ ($j=12,u$), are weights associated with the MSEs of gUE $u$. By setting the optimization variables as the equalizers and weights, then the relation between data rates and WMMSEs can be written by \cite{Joudeh2016}
\begin{equation}
\label{eq:wm1}
(\zeta_u^{12})^{\rm MMSE}= \min_{\mu_u^{12},f_u^{12}} \zeta_u^{12} =1-R_u^{12}\end{equation}
and 
\begin{equation}
\label{eq:wm2}
(\zeta_u^{u})^{\rm MMSE}= \min_{\mu_u^{u},f_u^{u}} \zeta_u^{u}=1-R_u^{u}.
\end{equation}
By substituting (\ref{eq:wm1})--(\ref{eq:wm2}) into (\ref{eq:awmse})--(\ref{eq:awmse2}), we obtain 
\begin{equation}
(\zeta_u^{12})^{\rm MMSE}= \mu_u^{12} (\varepsilon_u^{12})^{\rm MMSE} - \log_2(\mu_u^{12})\end{equation}
and
\begin{equation}
(\zeta_u^{u})^{\rm MMSE}= \mu_u^{u} (\varepsilon_u^{u})^{\rm MMSE} - \log_2(\mu_u^{u}),
\end{equation}
and the optimal MMSEs 
\begin{equation}
\label{eq:mu1}
(\mu_u^{12})^*=(\mu_u^{12})^{\rm MMSE}=\left((\varepsilon_u^{12})^{\rm MMSE}\right)^{-1}\end{equation}
and 
\begin{equation}
\label{eq:mu2}
(\mu_u^{u})^*=(\mu_u^{u})^{\rm MMSE}=\left((\varepsilon_u^{u})^{\rm MMSE}\right)^{-1}.
\end{equation}
}

\textcolor{black}{Motivated by the data rates-WMMSE relations in (\ref{eq:wm1})--(\ref{eq:wm2}), the optimization problem (P1) can be reformulated as
\begin{subequations}
	\begin{align}
	\small
	\min_{\mathbf{P}, \mathbf{x}, \mathbf{m}, \mathbf{f}} & \quad 
	\sum_{u=1}^2 v_u \zeta_{u}^{\rm tot} \tag{P2} \\
	\label{c1_P2}
	\text{s.t.}\quad & x_{1}^{12}+x_2^{12}+1\geq \zeta_{12},  \nonumber \tag{P2.a} \\
	\label{c2_P2} & \zeta_{u}^{\rm tot}\leq 1-R_{u}^{\rm{th}},\;\forall u \in \{1,2\} \tag{P2.b}\\
	\label{c3_P2} & \text{tr}(\mathbf{P}\mathbf{P}^H)\leq P_t  \tag{P2.c}\\
	\label{c4_P2} & \mathbf{x}\leq \mathbf{0}  \tag{P2.d} 
	\end{align}
\end{subequations}
where $\textbf{x}=[x_1^{12},x_2^{12}]=-\textbf{r}$, $\textbf{m}=[\mu_1^{12},\mu_2^{12},\mu_1^1,\mu_2^2]$, $\textbf{f}=[f_1^{12}, f_2^{12}, f_1^1,f_2^2]$, $\zeta_u^{\rm tot}=x_u^{12}+\zeta_u^u$ ($u=1,2$), and $\zeta_{12}=\max \{\zeta_1^{12},\zeta_2^{12}\}$.
}

\textcolor{black}{When minimizing the objective in (P2) for $\textbf{m}$ and $\textbf{f}$ (i.e., fixed $\textbf{P}$ and $\textbf{x}$), the optimal MMSE solution ($\textbf{m}^{\rm MMSE}$,$\textbf{f}^{\rm MMSE}$) is obtained 
according to (\ref{eq:equalizeropt})--(\ref{eq:equalizeropt2}) and (\ref{eq:mu1})--(\ref{eq:mu2}). These values satisfy the Karush Kuhn Tucker (KKT) optimality conditions in (P2) for $\textbf{P}$. Hence, given the relations in (\ref{eq:wm1})--(\ref{eq:wm2}) and the common rate transformation $\textbf{x}=-\textbf{r}$, (P2) can be transformed into (P1). Similarly, for any solution ($\textbf{P}^*$,$\textbf{x}^*$,$\textbf{m}^*$,$\textbf{f}^*$) that satisfies the KKT conditions in (P2), the solution ($\textbf{r}^*=-\textbf{x}^*$,$\textbf{P}^*$) satisfies the KKT conditions in (P1). Hence, (P1) can be transformed into (P2). However, (P2) is also non-convex for joint parameters optimization. To solve it, the authors proposed in \cite{Mao2018} to adopt an alternating optimization method. Indeed, in the $l^{th}$ iteration of the AO algorithm, the equalizers and weights are firstly updated using the precoders obtained in the $(l-1)^{th}$ iteration, i.e.,
\[
(\textbf{m},\textbf{f})=(\textbf{m}^{\rm MMSE}(\textbf{P}[l-1]),\textbf{f}^{\rm MMSE}(\textbf{P}[l-1])).
\]
Then, ($\textbf{x}$,$\textbf{P}$) is updated by solving (P2) for the given ($\mathbf{m}$,$\mathbf{f}$). Hence, ($\textbf{m}$, $\textbf{f}$) and ($\textbf{x}$, $\textbf{P}$) are iteratively updated until convergence of the WSR. The details of the procedure are presented in Algorithm \ref{Algo1}, where $[l]$ is the iteration index, $\mathbf{m}$ is the stream's weight vector, $\mathbf{f}$ is the equalizer vector, $\mathbf{x}$ is the transformation of $\mathbf{r}$, and $\delta \ll 1$ is the convergence condition \cite{Mao2018}.
}}


\textcolor{black}{Given Algorithm \ref{Algo1}, we can now define our placement iterative search method as follows: For each channel realization, $L$ locations at altitude $z_{aBS}$ are explored, where for each location RSMA parameters optimization is processed using Algorithm \ref{Algo1}. Then, the location resulting in the best WSR performance is selected. Algorithm \ref{Algo2} summarizes this process.
}

\begin{algorithm}[t]
\small{
\caption{Alternating Optimization Algorithm}
\label{Algo1}
\begin{algorithmic}[1]
\State {Initialize $l \xleftarrow{}0$, $\mathbf{P}[l]$, $\bar{R}[l]$} 
\Repeat
\State $l \xleftarrow{} l+1$; $\mathbf{P}[l-1] \xleftarrow{} \mathbf{P}$
\State $\mathbf{m} \xleftarrow{} \mathbf{m}(\mathbf{P}[l-1])$; $\mathbf{f} \xleftarrow{} \mathbf{f}(\mathbf{P}[l-1])$
\State Solve (P2) for updated ($\mathbf{m}, \mathbf{f}$), then update ($\mathbf{P}$, $\mathbf{x}$)
\Until {$|\bar{R}[l]-\bar{R}[l-1]|\leq \delta$}.
\end{algorithmic}}
\end{algorithm}

\begin{algorithm}[t]
\small{
\caption{3D Placement Iterative Search Algorithm}
\label{Algo2}
\begin{algorithmic}[1]
\State {Initialize matrix of aBS's 3D locations $\textbf{W}$ ($L \times 3$)} 
\State {Initialize vector of final WSRs $\bar{\textbf{R}}$ ($1 \times \Delta$)}
\State {Initialize vector of final best aBS locations ${\textbf{c}}$ ($1 \times \Delta$)}
\For {$l=1$ to $\Delta$} \% $\Delta:$ max. nbr. of channel realizations
\State Generate matrix $\textbf{H}$ ($L \times 2$) of channel coeff. (aBS to users) 
\For {$j=1$ to $L$}
\State {Initialize vector of WSRs for an aBS location  ${\textbf{R}}$ ($1 \times L$)}
\State Run Algo. \ref{Algo1} for aBS location $\textbf{W}[j,1:\text{end}]$
\State Get ${\textbf{R}}[j]$ from Algo. \ref{Algo1}
\EndFor
\State ${\bar{\textbf{R}}}[l]\xleftarrow{}\max_{j=1:L} \textbf{R}$ and $\textbf{c}[l]\xleftarrow{}\arg \max_{j=1:L} \textbf{R}$
\EndFor 
\State Return mean($\bar{\textbf{R}}$), $\bar{\textbf{R}}$ and $\textbf{c}$.
\end{algorithmic}}
\end{algorithm}

\subsubsection{Simulation Results}
We assume here that an aBS equipped with {$N=4$} antennas hovers at an altitude of {50 m}\footnote{{}{UAVs
should not fly above {}{122 m}, nor below {}{10 m}, due
to regulations and safety reasons \cite{FAA2019,Liu2018_2}.}} in order to serve {two} randomly spread gUEs in a $350 \times 300$ m$^2$ area. {For sake of simplicity, we assume that \textcolor{black}{$\sigma^2=1$}, $\alpha=2$, $v_1=v_2=\frac{1}{2}$, $B=1$ Hz, $R_1^{\rm th}=R_2^{\rm th}=0$ bps/Hz \cite{Mao2018}, $L=100$, $(a,b)=(9.61,0.16)$, and $(a_1,b_1)=(5, 15)$ dB (dense-urban environment) \cite{Azari2018,Bor2016}.}
The WSR performance is evaluated for different aBS placements and SNR values, given static users locations $[0, 0, 0]$ and $[100, 0, 0]$ m, and WSR is averaged over 100 channel realizations.

\begin{figure}[t]
\centering
\includegraphics[angle=-90,width=0.99\linewidth]{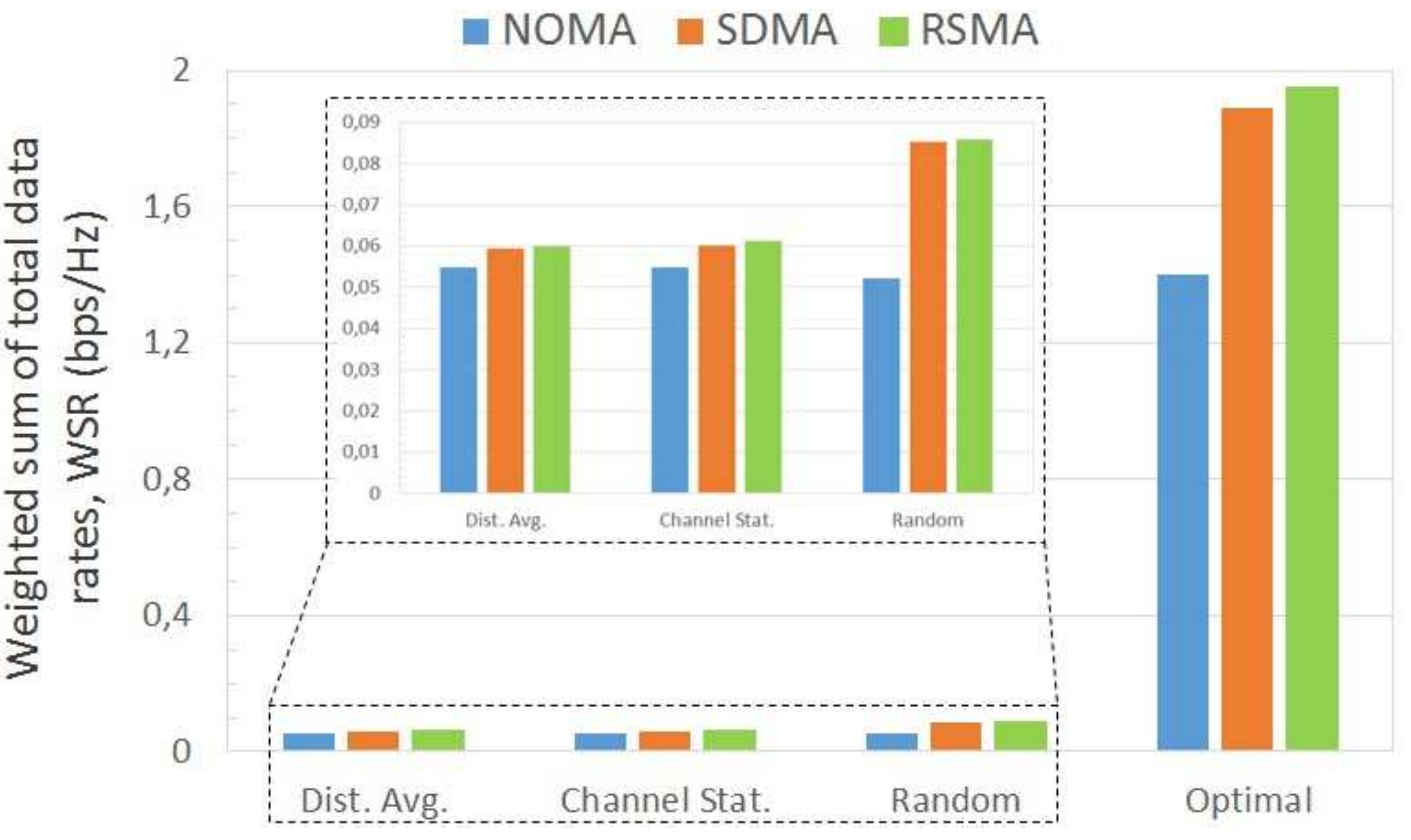}
\caption{WSR vs. aBS placement.}
\label{Fig1}
\end{figure}

Fig. \ref{Fig1} compares the WSR performance of NOMA, SDMA, and RSMA for different aBS placements, i.e., ``Dist. Avg." (Middle location between users at altitude $z_{aBS}$), ``Channel Stat." (Location obtained according to {\eqref{eq:optCoord}}), 
``Random" (location randomly selected \textcolor{black}{for each channel realization} over the segment linking the two users), and ``Optimal" (Location obtained using the proposed \textit{iterative search} method \textcolor{black}{for each channel realization}), given SNR=\textcolor{black}{$\frac{P_t}{\sigma^2}=
20$} dB. To be noted that the first two aBS placement approaches are static, while the last ones are dynamic (since they consider the small-scale channel realizations).
As shown, RSMA outperforms all other techniques in any aBS placement approach. Moreover, the ``Optimal" placement, taking into account the impact of instantaneous channel variations, provides the best performance. Compared to the static placement approaches, even ``Random" aBS placement achieves better performances.   

\textcolor{black}{\textit{Discussion:} Due to the time varying nature of the underlying wireless channels, it is clear that achieving the best aBS placement while optimizing the RSMA parameters is particularly complex because it requires a full and instantaneous knowledge of all channels in the investigated area. Such approach is unfeasible in a real system. Instead, an online implementation can be realized using an artificial intelligence agent mounted over the aBS, which is able, after a long enough training phase, to predict channel variations and decide on the best combination of aBS location and RSMA parameters in order to achieve the highest WSR performance. As the training phase tends to infinity, such method would converge to ``Optimal".}

\begin{figure}[t]
\centering
\includegraphics[width=0.95\linewidth]{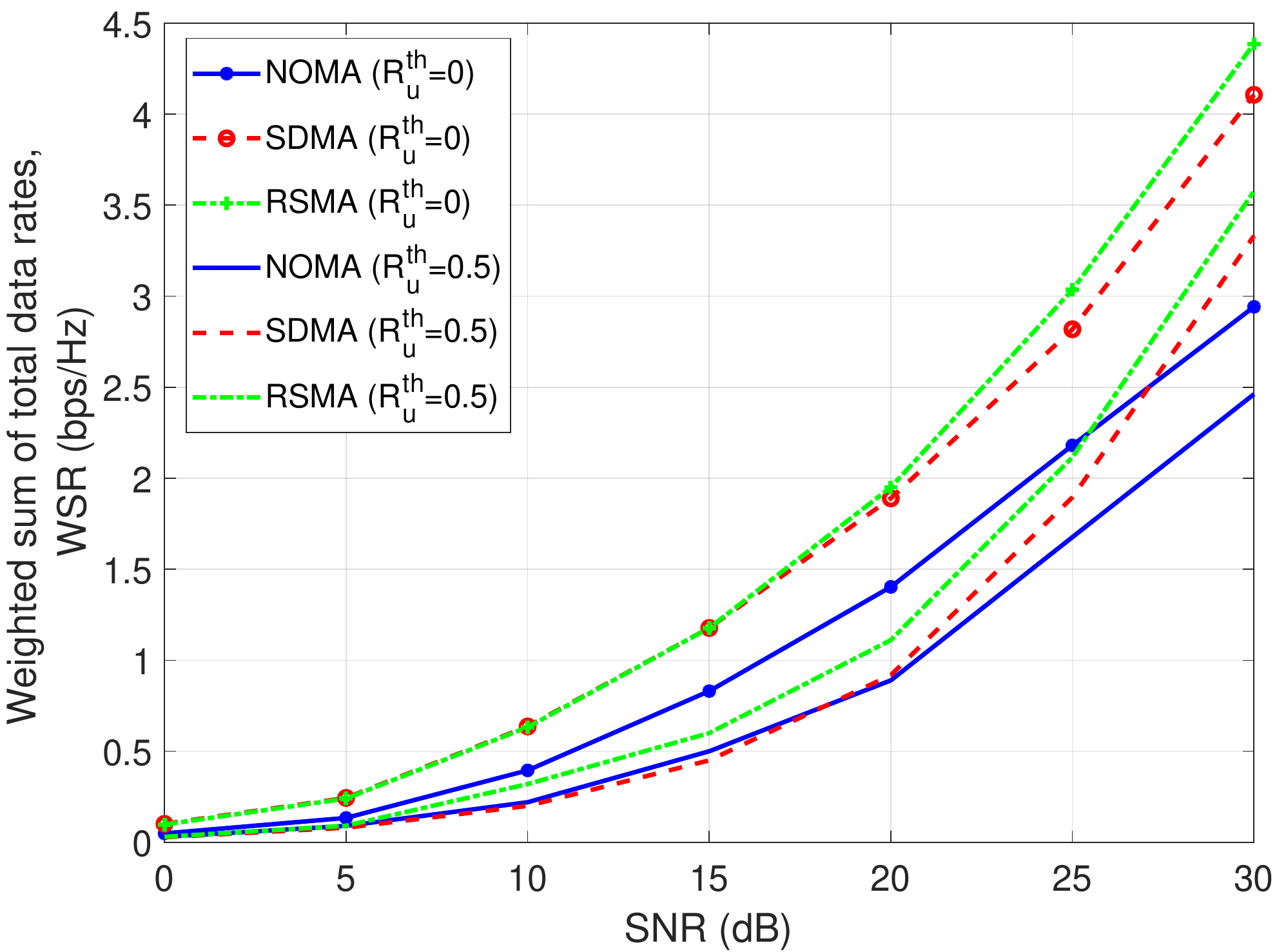}
\caption{WSR vs. SNR (``Optimal"), for different $R_u^{\rm th}$ ($u=1,2$).}
\label{Fig2}
\end{figure}

Fig. \ref{Fig2} depicts the  WSR performance as a function of the average SNR, \textcolor{black}{for different data rate thresholds $R_u^{\rm th}=0$ bps/Hz and $R_u^{\rm th}=0.5$ bps/Hz, respectively}. {For any $R_u^{\rm th}$ value, it} is shown that RSMA is superior to NOMA and SDMA, especially at high SNR. However, it should be pointed out here that RSMA exploits the advantages of flexible rate-splitting, SIC layers, and precoding, to allow for higher gains compared to NOMA and SDMA, where only SIC and/or precoding is used. {Nevertheless, SDMA performs better than NOMA at high SNR and for any $R_u^{\rm th}$, since the number of aBS antennas is greater than the number of served gUEs, which allows for better inter-user interference mitigation. However, it performs worse than NOMA 
at low SNR and $R_u^{\rm th}=0.5$ bps/Hz. This is mainly due to the fact that NOMA performs best for degraded channels and constrained data rates, while SDMA cannot efficiently alleviate interference in such conditions. Finally, for $R_u^{\rm th}=0.5$ bps/Hz, all access techniques perform worse than in the case of $R_u^{\rm th}=0$ bps/Hz. Indeed, as $R_u^{\rm th}>0$, constraint (P2.b) is leveraged, which restricts the feasible WMMSE solutions, hence degrading the overall WSR performance.}

\section{Open Issues and Research Directions}
This article presented the first look into the utilization of RSMA in aerial networks. We have shown that RSMA is a powerful multiple access scheme that can provide high data rates and reliable communications in aerial networks. However, several associated challenges need to be addressed and analyzed to realize the full potentials of RSMA.

\subsection{Joint Optimization of UAV and RSMA Parameters}
\textcolor{black}{The UAV provides additional degrees-of-freedom to the communication system through its deployment flexibility and mobility. Such properties can be exploited when optimizing the RSMA parameters, i.e., rate splitting, power allocation/beamforming design, and common messages design and ordering. In a dynamic environment in which users are mobile and/or communication channels are rapidly varying, a joint optimization of UAV and RSMA parameters would achieve high spectral-efficiency and energy-efficiency, with respect to the UAV's SWAP constraints. Such problems are typically NP-hard and require rigorous solutions that would rely on mixed-integer non-linear problem solving approaches, such as successive convex approximation, alternating optimization, and alternating direction method of multipliers, or smart methods based on artificial intelligence and machine learning.}

\subsection{Integration of RSMA into MIMO, Massive MIMO, and mmWave Aerial Systems}
As discussed previously, studies on MIMO and massive MIMO multiple access in aerial networks are either limited or absent in the literature. Indeed, most existing work has focused on a single setup where a single aBS serves multiple users using a single frequency band, e.g., NOMA. However, with the recent emergence of multiple access in multi-carrier systems as a study area, as in \cite{Zeng2018_2}, an investigation into performance improvements with MIMO, or massive-MIMO with multi-carrier systems for NOMA, SDMA, and RSMA would be a promising research direction.

Yet, most work has focused on downlink transmission, with only a few taking interest in uplink. For instance, \cite{Zeng2019_n} proposed SIMO-NOMA for the uplink of an IoT system, where a rate-splitting scheme that guaranteed max-min fairness was studied. Authors in \cite{Hu2019_n} applied NOMA to the uplink of a relay-assisted wireless sensor network, whereas \cite{Liu2019_n} proposed enhancing the outage performance of an uplink NOMA system using rate splitting. None of the aforementioned approaches were applied to aerial networks, and all of them focused solely on NOMA. Hence, the investigation of uplink MIMO-RSMA in the context of aerial networks is an unexplored topic.

Coexistence of mmWave and RSMA systems is partially addressed in \cite{Rahmati2019}, where the gBS relies on mmWave to serve aerial users. Nevertheless, using mmWave-enabled UAVs can offer many benefits. For instance, high-bandwidth aBSs, data collectors, and even aerial mobile edge computing servers or caching devices can be deployed to achieve on-the-fly access. Being superior to NOMA and SDMA, RSMA in the aforementioned scenarios will bring substantial performance improvements.

\subsection{Distributed and Cooperative RSMA in Hybrid Terrestrial/Non-terrestrial Networks}

Research on integrated terrestrial/non-terrestrial networks has recently been increasing \cite{Liu2018_sagin}. One of the main challenges to be addressed in such systems is traffic offloading. Indeed, unlike terrestrial networks, where traffic offloading approaches are mature, moving traffic from terrestrial to satellite and/or UAV
networks is still in its infancy, and challenges of propagation delay, link selection, and channel allocation remain. In contexts like this, RSMA can be introduced in the integrated network, either horizontally (at one of the terrestrial or non-terrestrial layers) or vertically (through cooperative heterogeneous layers) in order to improve the quality of communications. RSMA transmission management can be then realized centrally, assuming that a central controller manages each layer, or distributively, given than layers are separately managed either horizontally or vertically. The investigation of joint optimization problems in such complex systems has not been carried out yet. 
 
Moreover, exploring distributed RSMA using UAV swarms may be of interest in critical scenarios, where reliable communication is required. The use of UAV swarms with RSMA is challenging, as problems of UAVs placements/trajectories and RSMA parameters optimization have to be jointly addressed. 

Finally, RSMA can be analyzed in different aerial network topologies, e.g., UAVs may act as relays in a multi-hop or mesh topology, where joint UAVs placement, RSMA precoding, and rate-splitting issues are worth investigating to achieve objectives, such as {spectral efficiency} and {energy efficiency} maximization.

{
\subsection{Physical Layer Security Issues in Aerial Multiple Access Networks}
Recently, wireless communication researchers started focusing increasingly on the physical layer security of emerging wireless technologies. Indeed, the latter can efficiently complement the conventional cryptography-based methods to enhance the security of wireless systems. Within the context of aerial networks, this issue is even more challenging due to the open-sky wireless environment. Not only is it likely for data transmission to be intercepted by an eavesdropper, but also it is possible for UAVs to experience malicious jamming attacks. Consequently, there is an urgent need to prevent attacks, such as eavesdropping, man-in-the-middle attack, denial-of-service, etc. From the UAV's perspective, several effective solutions have been proposed. For instance, multiple access techniques can be leveraged to communicate and generate artificial noise (jamming signals) towards external eavesdroppers, and hence, blocking them from decoding useful signals. Such method was investigated for SDMA and NOMA in \cite{furqan2019physical}, but no extension to RSMA has been realized yet. Moreover, in order to avoid data leakage to an internal eavesdropper when SIC receivers are used, users' data can be protected either by using cryptography methods, or through data transformation to different domains, e.g., using channel-dependent features, which minimizes data's decodability by different users than the intended one. However, current contributions in this direction are still limited. Finally, blockchain technology can be leveraged to secure transiting data through the UAVs. These topics will be investigated extensively in future relevant contributions.   
}

\section {Conclusion}
In this paper, we reviewed orthogonal and non-orthogonal multiple access techniques for aerial networks and showed their advantages and limitations. Then, we surveyed most recent RSMA works and highlighted the limited use of RSMA in aerial systems. Subsequently, we discussed the integration of RSMA into aerial networks, where results for a simple integration case were presented. Both aBS placement and RSMA parameters were optimized to maximize the weighted sum rate of users. 
It was shown that in this system, RSMA outperforms both NOMA and SDMA due to its rate-splitting flexibility and precoding efficiency. Finally, a number of open issues and research directions, linked to the application of RSMA in aerial networks, have been presented and discussed.  

\balance
\bibliographystyle{IEEEtran}
\bibliography{IEEEabrv,tau}

\end{document}